\documentclass[english,onecolumn]{IEEEtran}
\usepackage[T1]{fontenc}
\usepackage[latin9]{inputenc}
\usepackage{color}
\usepackage{array}
\usepackage{mathrsfs}
\usepackage{multirow}
\usepackage{amsmath}
\usepackage{amsthm}
\usepackage{amssymb}
\usepackage{graphicx}
\usepackage{psfrag}
\usepackage{esint}

\makeatletter

\theoremstyle{plain}
\newtheorem{thm}{\protect\theoremname}
\theoremstyle{plain}
\newtheorem{prop}{\protect\propositionname}
\theoremstyle{plain}

\usepackage{xspace}
\usepackage{bbm}
\input{mathlig}

\newcommand{\muspace}{\mspace{1mu}}

\DeclareRobustCommand{\scond}{\mathchoice{\muspace\vert\muspace}{\vert}{\vert}{\vert}}
\mathlig{|}{\scond}

\DeclareRobustCommand{\discint}{\mathchoice{\mspace{-1.5mu}:\mspace{-1.5mu}}{\mspace{-1.5mu}:\mspace{-1.5mu}}{:}{:}}
\mathlig{::}{\discint}

\newcommand{\Xc}{\mathcal{X}}
\newcommand{\Yc}{\mathcal{Y}}

\newcommand{\Rr}{\mathscr{R}}

\newcommand{\Xh}{{\hat{X}}}
\newcommand{\Yh}{{\hat{Y}}}

\DeclareMathOperator\E{\textsf{E}}
\let\P\relax
\DeclareMathOperator\P{\textsf{P}}

\def\textiid{i.i.d.\@\xspace}
\newcommand\iid{\ifmmode\text{ i.i.d. } \else \textiid \fi}

\def\mathllap{\mathpalette\mathllapinternal}
\def\mathllapinternal#1#2{%
  \llap{$\mathsurround=0pt#1{#2}$}}

\def\clap#1{\hbox to 0pt{\hss#1\hss}}
\def\mathclap{\mathpalette\mathclapinternal}
\def\mathclapinternal#1#2{%
  \clap{$\mathsurround=0pt#1{#2}$}}

\let\oldstackrel\stackrel
\renewcommand{\stackrel}[2]{\oldstackrel{\mathclap{#1}}{#2}}

\renewcommand{\hbar}{h\mathllap{\overline{\vphantom{h}\hphantom{\rule{4.6pt}{0pt}}}\mspace{0.77mu}}}

\catcode`~=11 
\newcommand{\urltilde}{\kern -.06em\lower -.06em\hbox{~}\kern .02em}
\catcode`~=13 

\renewcommand{\E}{\mbox{\normalfont\textsf{E}}}
\renewcommand{\P}{\mbox{\normalfont\textsf{P}}}

\usepackage{mathrsfs}
\usepackage{colortbl}
\definecolor{lightred}{rgb}{1,0.6,0.6}
\definecolor{lightgreen}{rgb}{0.6,1,0.6}
\definecolor{lightyellow}{rgb}{1,1,0.5}
\definecolor{lightgrey}{rgb}{0.8,0.8,0.8}

\makeatother

\usepackage{babel}
\providecommand{\corollaryname}{Corollary}
\providecommand{\propositionname}{Proposition}
\providecommand{\theoremname}{Theorem}

\begin{document}

\title{Extended Gray--Wyner System with Complementary Causal Side Information}

\author{Cheuk Ting Li and Abbas El Gamal\\
 Department of Electrical Engineering, 
 Stanford University\\
 Email: ctli@stanford.edu, abbas@ee.stanford.edu 
}
\maketitle
\begin{abstract}
We establish the rate region of an extended Gray--Wyner system for 2-DMS $(X,Y)$ with two additional decoders having complementary causal side information. This extension is interesting because in addition to the operationally significant extreme points of the Gray--Wyner rate region, which include Wyner's common information, G{\'a}cs-K{\"o}rner common information and information bottleneck, the rate region for the extended system also includes the  K{\"o}rner graph entropy, the privacy funnel and excess functional information, as well as three new quantities of potential interest, as extreme points. To simplify the investigation of the 5-dimensional rate region of the extended Gray--Wyner system, we establish an equivalence of this region to a 3-dimensional mutual information region that consists of the set of all triples of the form $(I(X;U),\,I(Y;U),\,I(X,Y;U))$ for some $p_{U|X,Y}$. We further show that projections of this mutual information region yield the rate regions for many settings involving a 2-DMS, including lossless source coding with causal side information, distributed channel synthesis, and lossless source coding with a helper.
\end{abstract}

\section{Introduction}


The lossless Gray--Wyner system~\cite{Gray--Wyner1974} is a multi-terminal source coding setting for two discrete memoryless source (2-DMS) $(X,Y)$ with one encoder and two decoders. This setup draws some of its significance from providing operational interpretation for several information theoretic quantities of interest, namely Wyner's common information~\cite{Wyner1975a}, the G{\'a}cs-K{\"o}rner common information~\cite{Gacs--Korner1973}, the necessary conditional entropy~\cite{cuff2010coordination}, and the information bottleneck~\cite{tishby2000information}.

\begin{figure}[h]
\begin{center}
\psfrag{xy}[b]{$X^n,Y^n$}
\psfrag{x3}[l]{$Y^i$}
\psfrag{y4}[l]{$X^i$}
\psfrag{rw}[c]{Encoder}
\psfrag{d1}[c]{Decoder 1}
\psfrag{d2}[c]{Decoder 2}
\psfrag{d3}[c]{Decoder 3}
\psfrag{d4}[c]{Decoder 4}
\psfrag{xh}[b]{$\Xh_1^n$}
\psfrag{yh}[b]{$\Yh_2^n$}
\psfrag{xh3}[b]{$\Xh_{3i}$}
\psfrag{yh4}[b]{$\Yh_{4i}$}
\psfrag{ix}[b]{$M_1$}
\psfrag{iy}[b]{$M_2$}
\psfrag{j}[b]{$M_0$}
\psfrag{ix3}[b]{$M_3$}
\psfrag{iy4}[b]{$M_4$}
\includegraphics[scale=0.55]{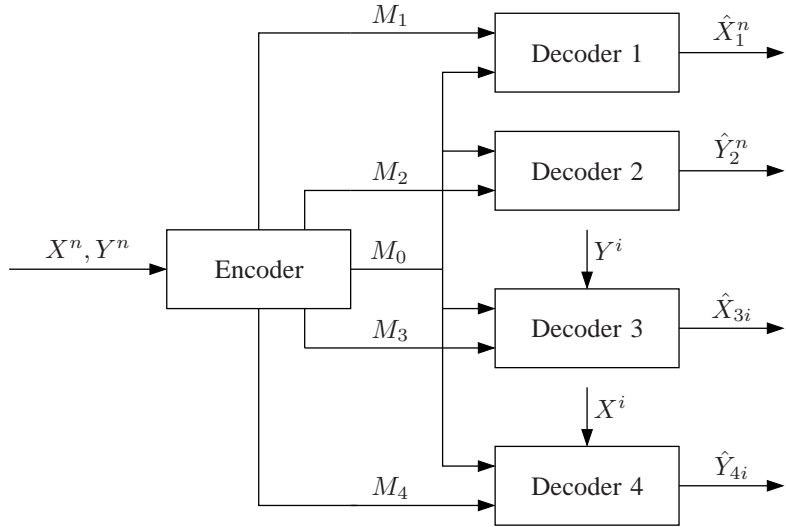}
\caption{\label{fig:gw}Extended Gray--Wyner system.}
\end{center}
\end{figure}

In this paper, we consider an extension of the Gray-Wyner system (henceforth called the EGW system), which includes two new individual descriptions and two decoders with causal side information as depicted in Figure~\ref{fig:gw}. The encoder maps sequences from a 2-DMS $(X,Y)$ into five indices $M_{i}\in [1:2^{nR_i}]$, $i=0,\ldots,4$. Decoders 1 and 2 correspond to those of the Gray--Wyner system, that is, decoder 1 recovers $X^n$ from $(M_0,M_1)$ and decoder 2 recovers $Y^n$ from $(M_0,M_2)$. At time $i\in [1:n]$,  decoder 3 recovers $X_i$ \emph{causally} from $(M_0,M_3, Y^i)$ and decoder 4 similarly recovers $Y_i$ causally from $(M_0,M_4,X^i)$. Note that decoders 3 and 4 correspond to those of the complementary delivery setup studied in~\cite{wyner2002satellite,Kimura--Uyematsu2006} with causal (instead of noncausal) side information and with two additional private indices $M_3$ and $M_4$. This extended Gray-Wyner system setup is lossless, that is, the decoders recover their respective source sequences with probability of error that vanishes as $n$ approaches infinity. The rate region $\Rr$ of the EGW system is defined in the usual way as the closure of the set of achievable rate tuples $(R_0,R_1,R_2,R_3,R_4)$.

The first contribution of this paper is to establish the rate region of the EGW system. Moreover, to simplify the study of this rate region and its extreme points, we show that it is equivalent to the 3-dimensional \emph{mutual information region} for $(X,Y)$ defined as
\begin{align}
\mathscr{I}_{XY} &=\bigcup_{p_{U|XY}}\left\{ (I(X;U),\,I(Y;U),\,I(X,Y;U))\right\} \subseteq\mathbb{R}^{3} \label{eq:IXY}
\end{align}
in the sense that we can express $\Rr$ using $\mathscr{I}$ and vice versa. As a consequence and of particular interest, the extreme points of the rate region $\Rr$ (and its equivalent mutual information region $\mathscr{I}_{XY}$) for the EGW system include, in addition to the aforementioned extreme points of the Gray--Wyner system, the K{\"o}rner graph entropy~\cite{korner1973coding}, privacy funnel~\cite{makhdoumi2014funnel} and excess functional information~\cite{sfrl_arxiv}, as well as three new quantities with interesting operational meaning, which we refer to as the {\em maximal interaction information}, the {\em asymmetric private interaction information}, and the {\em symmetric private interaction information}. These extreme points can be cast as maximizations of the interaction information~\cite{mcgill1954multivariate} $I(X;Y|U)-I(X;Y)$ under various constraints. They can be considered as distances from extreme dependency, as they are equal to zero only under certain conditions of extreme dependency.  In addition to providing operational interpretations to these information theoretic quantities, projections of the mutual information region yield the rate regions for many settings involving a 2-DMS, including lossless source coding with causal side information~\cite{Weissman--El-Gamal2006}, distributed channel synthesis~\cite{cuff2013distributed,bennet2014reverse}, and lossless source coding with a helper~\cite{Wyner1973b,Ahlswede--Korner1975,Wyner1975b}. 

A related extension of lossy Gray--Wyner system with two decoders with causal side information was studied
by Timo and Vellambi~\cite{timo2009causal}. If we only consider decoders 3 and 4 in EGW, then it can be considered as a special case of their setting (where the side information does not need to be complementary). Other related source coding setups to the EGW can be found in~\cite{Wyner--Ziv1976,Weissman--El-Gamal2006,heegard1985sideinfo,steinberg2004successive,tian2008sideinfo}. A related 3-dimensional region, called the region of tension, was investigated by Prabhakaran and Prabhakaran~\cite{prabhakaran2014assisted,prabhakaran2014tension}. 
We show that this region can be obtained from the mutual information region, but the other direction does not hold in general.

In the following section, we establish the rate region of the EGW system, relate it to the mutual information region, and show that the region of the original Gray--Wyner system and the region of tension can be obtained from the mutual information region. In Section~\ref{sec:extremepts}, we study the extreme points of the mutual information region. In Section~\ref{sec:asymp} we establish the rate region for the same setup as the EGW system but with noncausal instead of causal side information at decoders 3 and 4. We show that the rate region of the noncausal EGW can be expressed in terms of the Gray--Wyner region, hence it does not contain as many interesting extreme points as the causal EGW. Moreover, we show that this region is equivalent to the closure of the limit of the mutual information region for $(X^n,Y^n)$ as $n$ approaches infinity.

\subsection{Notation}

Throughout this paper, we assume that $\log$ is base 2 and the entropy
$H$ is in bits. We use
the notation: $X_{a}^{b}=(X_{a},\ldots,X_{b})$, $X^{n}=X_{1}^{n}$ and
$[a:b]=[a,b]\cap\mathbb{Z}$.

For discrete $X$, we write the probability mass function
as $p_{X}$. For $A\subseteq\mathbb{R}^{n}$, we write the closure of $A$ as $\mathrm{cl}(A)$
and the convex hull as $\mathrm{conv}(A)$. We write the support function
as
\[
\psi_{A}(b)=\sup\left\{ a^{T}b:\,a\in A\right\} .
\]
We write the one-sided directional derivative of the support function
as
\[
\psi'_{A}(b;c)=\lim_{t\to0^{+}}\frac{1}{t}\left(\psi_{A}(b+tc)-\psi_{A}(b)\right).
\]
Note that if $A$ is compact and convex, then
\[
\psi'_{A}(b;c)=\max\left\{ d^{T}c:\,d\in\underset{a\in A}{\arg\max}\,a^{T}b\right\}.
\]

\section{Rate region of EGW and the mutual information region\label{sec:asymp-1}}

The rate region of the EGW system is given in the following.
\begin{thm}
\label{thm:gwc}The rate region the EGW system $\mathscr{R}$
is the set of rate tuples $(R_{0},R_1,R_2,R_3,R_{4})$ such that
\begin{align*}
R_{0} & \ge I(X,Y;U),\\
R_{1} & \ge H(X|U),\\
R_{2} & \ge H(Y|U),\\
R_{3} & \ge H(X|Y,U),\\
R_{4} & \ge H(Y|X,U)
\end{align*}
for some $p_{U|XY}$, where $|\mathcal{U}|\le|\mathcal{X}|\cdot|\mathcal{Y}|+2$.
\end{thm}
Note that if we ignore decoders 3 and 4, i.e., let $R_{3},R_{4}$ be sufficiently
large, then this region reduces to the Gray--Wyner region.
\begin{IEEEproof}
The converse proof is quite straightforward and is given in Appendix~\ref{subsec:pf_gwc} for completion.
We now prove the achievability. 

\noindent \emph{Codebook generation.} Fix $p_{U|XY}$ and randomly and independently generate $2^{nR_{0}}$
sequences $u^{n}(m_{0})$, $m_{0}\in[1:2^{nR_{0}}]$, each according
to $\prod_{i=1}^{n}p_{U}(u_{i})$. Given $u^{n}(m_{0})$, assign indices
$m_{1}\in[1:2^{nR_{1}}]$, $m_{2}\in[1:2^{nR_{2}}]$ to the sequences
in the conditional typical sets $\mathcal{T}_{\epsilon}^{(n)}(X|u^{n}(m_{0}))$
and $\mathcal{T}_{\epsilon}^{(n)}(Y|u^{n}(m_{0}))$, respectively.
For each $y\in\mathcal{Y}$, $u\in\mathcal{U}$, assign indices $m_{3,y,u}\in[1:2^{nR_{3,y,u}p_{YU}(y,u)}]$
to the sequences in $\mathcal{T}_{\epsilon}^{n(1+\epsilon)p_{YU}(y,u)}(X|y,u)$,
where $\sum_{y,u}R_{3,y,u}p_{YU}(y,u)\le R_{3}$. Define $m_{4,x,u}$
similarly.
\smallskip

\noindent \emph{Encoding.} To encode the sequence $x^{n},y^{n}$, find $m_{0}$ such that $(u^{n}(m_{0}),x^{n},y^{n})\in\mathcal{T}_{\epsilon}^{(n)}$ is jointly typical, and find indices $m_{1},m_{2}$ of $x^{n},y^{n}$
in $\mathcal{T}_{\epsilon}^{(n)}(X|u^{n}(m_{0}))$ and $\mathcal{T}_{\epsilon}^{(n)}(Y|u^{n}(m_{0}))$
given $u^{n}(m_{0})$. For each $x,y$, let $x_{y,u}^{n}$ be the
subsequence of $x^{n}$ where $x_{i}$ is included if and only if
$y_{i}=y$ and $u_{i}(m_{0})=u$. Note that since $(u^{n}(m_{0}),y^{n})\in\mathcal{T}_{\epsilon}^{(n)}$,
the length of $x_{y,u}^{n}$ is not greater than $n(1+\epsilon)p_{YU}(y,u)$.
We then find an index $m_{3,y,u}$ of $\hat{x}_{y,u}^{n(1+\epsilon)p_{YU}(y,u)}\in\mathcal{T}_{\epsilon}^{n(1+\epsilon)p_{YU}(y,u)}(X|y,u)$
such that $x_{y,u}^{n}$ is a prefix of $\hat{x}_{y,u}^{n(1+\epsilon)p_{YU}(y,u)}$,
and output $m_{3}$ as the concatenation of $m_{3,y,u}$ for all $y,u$.
Similar for $m_{4}$.
\smallskip

\noindent \emph{Decoding.} Decoder 1 outputs the sequence corresponding to the index $m_{1}$ in $\mathcal{T}_{\epsilon}^{(n)}(X|u^{n}(m_{0}))$. Decoder 2 performs similarly using $(m_0,m_2)$. Decoder 3, upon observing $y_{i}$, finds the
sequence $\hat{x}_{y_{i},u_{i}(m_{0})}^{n(1+\epsilon)p_{YU}(y_{i},u_{i}(m_{0}))}$
at the index $m_{3,y_{i},u_{i}(m_{0})}$ in $\mathcal{T}_{\epsilon}^{n(1+\epsilon)p_{YU}(y_{i},u_{i}(m_{0}))}(X|y_{i},u_{i}(m_{0}))$,
and output the next symbol in the sequence that is not previously used. Decoder 4 performs similarly using $(m_0,m_4)$.
\smallskip 

\noindent \emph{Analysis of the probability of error.} By the covering lemma, the probability that there does not exist $m_{0}$
such that $(u^{n}(m_{0}),x^{n},y^{n})\in\mathcal{T}_{\epsilon}^{(n)}$
tends to 0 if $R_{0}>I(X,Y;U)$. Also $\big|\mathcal{T}_{\epsilon}^{(n)}(X|u^{n}(m_{0}))\big|\le2^{nR_{1}}$
for large $n$ if $R_{1}>H(X|U)+\delta(\epsilon)$ (similar for $R_{2}>H(Y|U)+\delta(\epsilon)$).
Note that $(u^{n}(m_{0}),x^{n},y^{n})\in\mathcal{T}_{\epsilon}^{(n)}$
implies
\begin{align*}
\frac{\left|\left\{ i:\,x_{i}=x,\,y_{i}=y,\,u_{i}(m_{0})=u\right\} \right|}{n(1+\epsilon)p_{YU}(y,u)} & \le\frac{(1+\epsilon)p_{XYU}(x,y,u)}{(1+\epsilon)p_{YU}(y,u)}\\
 & \le p_{X|YU}(x|y,u)
\end{align*}
for all $(y,u)$. Hence there exists $\hat{x}_{y,u}^{n(1+\epsilon)p_{YU}(y,u)}\in\mathcal{T}_{\epsilon}^{n(1+\epsilon)p_{YU}(y,u)}(X|y,u)$
such that $x_{y,u}^{n}$ is a prefix of $\hat{x}_{y,u}^{n(1+\epsilon)p_{YU}(y,u)}$.
And $\big|\mathcal{T}_{\epsilon}^{n(1+\epsilon)p_{YU}(y,u)}(X|y,u)\big|\le2^{nR_{3,y,u}p_{YU}(y,u)}$
for large $n$ if $R_{3,y,u}>(1+\epsilon)H(X|Y=y,U=u)+\delta(\epsilon)$.
Hence we can assign suitable $R_{3,y,u}$ for each $y,u$ if $R_{3}>(1+\epsilon)H(X|Y,U)+\delta(\epsilon)$.

\end{IEEEproof}


Although $\mathscr{R}$ is 5-dimensional, the bounds on the rates can be expressed in terms of three quantities: $I(X;U)$, $I(Y;U)$ and $I(X,Y;U)$ together with other constant quantities that involve only the given $(X,Y)$. This leads to the following equivalence of $\Rr$ to the mutual information region $\mathscr{I}_{XY}$ defined in~\eqref{eq:IXY}. We denote the components of a vector $v\in\mathscr{I}_{XY}$ by $v=(v_{X},\,v_{Y},\,v_{XY})$.

\begin{prop}
\label{prop:equiv}The rate region for the EGW system can be expressed as
\begin{align}
\mathscr{R}=\bigcup_{v\in\mathscr{I}_{XY}} & \big\{ \big(v_{XY},\,H(X)-v_{X},\,H(Y)-v_{Y},\,H(X|Y)-v_{XY}+v_{Y},\,H(Y|X)-v_{XY}+v_{X}\big) \big\}+[0,\infty)^5,\label{eq:gwc_ixy}
\end{align}
where the last ``$+$'' denotes the Minkowski sum. Moreover, the mutual information region for $(X,Y)$ can be expressed as
\begin{align}
\mathscr{I}_{XY}=\left\{ v \in \mathbb{R}^3 :\,\big(v_{XY},\,H(X)-v_{X},\,H(Y)-v_{Y},\,H(X|Y)-v_{XY}+v_{Y},\,H(Y|X)-v_{XY}+v_{X}\big)\in\Rr\right\}.\label{eq:ixy_gwc}
\end{align}
\end{prop}
\begin{IEEEproof}
Note that~\eqref{eq:gwc_ixy} follows from the definitions of $\Rr$ and $\mathscr{I}_{XY}$. We now prove~\eqref{eq:ixy_gwc}. The $\subseteq$ direction follows from~\eqref{eq:gwc_ixy}. For the $\supseteq$ direction, let $v \in \mathbb{R}^3$ satisfy
\[
\big(v_{XY},\,H(X)-v_{X},\,H(Y)-v_{Y},\,H(X|Y)-v_{XY}+v_{Y},\,H(Y|X)-v_{XY}+v_{X}\big)\in\mathscr{R}.
\]
Then by Theorem~\ref{thm:gwc}, there exists $U$ such that
\begin{gather}
v_{XY}  \ge I(X,Y;U),\label{eq:equiv_pf_a}\\
H(X)-v_{X} \ge H(X|U),\label{eq:equiv_pf_b}\\
H(Y)-v_{Y} \ge H(Y|U),\label{eq:equiv_pf_c}\\
H(X|Y)-v_{XY}+v_{Y} \ge H(X|Y,U),\label{eq:equiv_pf_d}\\
H(Y|X)-v_{XY}+v_{X} \ge H(Y|X,U).\label{eq:equiv_pf_e}
\end{gather}
Adding~\eqref{eq:equiv_pf_a} and~\eqref{eq:equiv_pf_e}, we have $v_X \ge I(X;U)$. Combining this with~\eqref{eq:equiv_pf_b}, we have $v_X = I(X;U)$. Similarly $v_Y = I(Y;U)$. Substituting this into~\eqref{eq:equiv_pf_d}, we have $v_{XY} \le I(X,Y;U)$. Combining this with~\eqref{eq:equiv_pf_a}, we have $v_{XY} = I(X,Y;U)$. Hence $v \in \mathscr{I}_{XY}$.
\end{IEEEproof}

In the following we list several properties of $\mathscr{I}_{XY}$. 
\begin{prop}
\label{prop:ixy_prop}The mutual information region $\mathscr{I}_{XY}$
satisfies:
\begin{enumerate}
\item Compactness and convexity. $\mathscr{I}_{XY}$ is compact and convex.
\item Outer bound. $\mathscr{I}_{XY}\subseteq\mathscr{I}^{\mathrm{o}}_{XY}$,
where $\mathscr{I}^{\mathrm{o}}_{XY}$ is the set of $v$ such that
\begin{gather*}
v_{X},\,v_{Y}\ge0,\\
v_{X}+v_{Y}-v_{XY}\le I(X;Y),\\
0\le v_{XY}-v_{Y}\le H(X|Y),\\
0\le v_{XY}-v_{X}\le H(Y|X).
\end{gather*}
\item Inner bound. $\mathscr{I}_{XY}\supseteq\mathscr{I}^{\mathrm{i}}_{XY}$,
where $\mathscr{I}^{\mathrm{i}}_{XY}$ is the convex hull of the points $(0,\,0,\,0), (H(X),\,I(X;Y),\,H(X)), \\(I(X;Y),\,H(Y),\,H(Y)), (H(X),\,H(Y),\,H(X,Y)), (H(X|Y),\,H(Y|X),\,H(X|Y)+H(Y|X))$.

Moreover, there exists $0\le\epsilon_{1},\epsilon_{2}\le\log I(X;Y)+4$ such that
\[
(0,\,H(Y|X)-\epsilon_{1},\,H(Y|X)), \, (H(X|Y)-\epsilon_{2},\,0,\,H(X|Y)) \in \mathscr{I}_{XY}.
\]
\item Superadditivity. If $(X_{1},Y_{1})$ is independent of $(X_{2},Y_{2})$,
then
\[
\mathscr{I}_{X_{1},Y_{1}}+\mathscr{I}_{X_{2},Y_{2}}\subseteq\mathscr{I}_{(X_{1},X_{2}),(Y_{1},Y_{2})},
\]
where $+$ denotes the Minkowski sum. As a result, if $(X_{i},Y_{i})\sim p_{XY}$
i.i.d. for $i=1,\ldots,n$,
$\mathscr{I}_{XY}\subseteq (1/n)\mathscr{I}_{X^{n},Y^{n}}$.
\item Data processing. If $X_{2}-X_{1}-Y_{1}-Y_{2}$ forms a Markov chain,
then for any $v\in\mathscr{I}_{X_{1},Y_{1}}$, there exists $w\in\mathscr{I}_{X_{2},Y_{2}}$
such that
$w_{X}\le v_{X},\;w_{Y}\le v_{Y},\;w_{XY}\le v_{XY},$
\[
I(X_{2};Y_{2})-w_{X}-w_{Y}+w_{XY}\le I(X_{1};Y_{1})-v_{X}-v_{Y}+v_{XY}.
\]
\item Cardinality bound.
\[
\mathscr{I}_{XY}=\bigcup_{p_{U|XY}:\,|\mathcal{U}|\le|\mathcal{X}|\cdot|\mathcal{Y}|+2}\left\{ (I(X;U),\,I(Y;U),\,I(X,Y;U))\right\} .
\]
\item Relation to Gray--Wyner region and region of tension. The Gray--Wyner region can be obtained from $\mathscr{I}_{XY}$ as
\begin{align*}
\mathscr{R}_{\mathrm{GW}} &=\bigcup_{p_{U|XY}} \big\{ \big(I(X,Y;U),\,H(X|U),\,H(Y|U) \big) \big\} + [0,\infty)^3 \\
& = \bigcup_{v\in\mathscr{I}_{XY}}  \big\{ \big(v_{XY},\,H(X)-v_{X},\,H(Y)-v_{Y}\big) \big\}+[0,\infty)^3.
\end{align*}
The region of tension can be obtained from $\mathscr{I}_{XY}$ as
\begin{align*}
\mathfrak{T} &=\bigcup_{p_{U|XY}} \big\{ \big(I(Y;U|X),\,I(X;U|Y),\,I(X;Y|U) \big) \big\}+ [0,\infty)^3 \\
& = \bigcup_{v\in\mathscr{I}_{XY}}  \big\{ \big(v_{XY} - v_X,\,v_{XY} - v_Y,\,I(X;Y) - v_X - v_Y + v_{XY}\big) \big\}+[0,\infty)^3.
\end{align*}
\end{enumerate}
\end{prop}
The proof of this proposition is given in Appendix~\ref{subsec:pf_ixy_prop}.

\section{Extreme Points of the Mutual Information Region\label{sec:extremepts}}


Many interesting information theoretic quantities can be expressed as optimizations over $\mathscr{I}_{XY}$ (and $\mathscr{R}$).
Since $\mathscr{I}_{XY}$ is convex and compact,
some of these quantities can be represented in terms
of the support function $\psi_{\mathscr{I}_{XY}}(x)$ and its one-sided
directional derivative, which provides a representation of those quantities using at most 6 coordinates. To avoid conflicts and for consistency, we use different notation for some of these quantities from the original literature . We use semicolons, e.g., $G(X;Y)$, for symmetric quantities, and arrows, e.g., $G(X\to Y)$, for
asymmetric quantities.

\begin{figure}[h]
\begin{centering}
\includegraphics[scale=0.75]{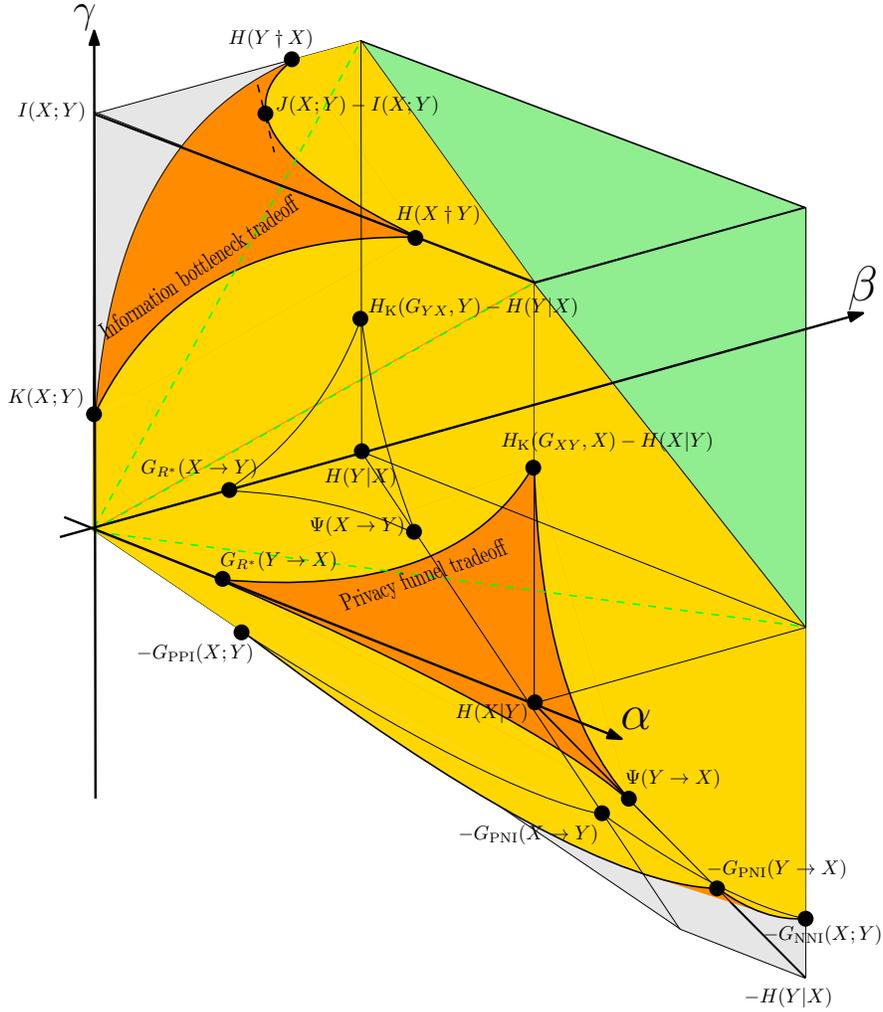}
\par\end{centering}
\caption{\label{fig:ixy_3d}Illustration of $\mathscr{I}_{XY}$ (yellow), $\mathscr{I}^{\mathrm{i}}_{XY}$
(green) and $\mathscr{I}^{\mathrm{o}}_{XY}$ (grey) defined in Proposition~\ref{prop:ixy_prop}. The axes are $\alpha=I(X;U|Y)=v_{XY}-v_{Y}$,
$\beta=I(Y;U|X)=v_{XY}-v_{X}$ and $\gamma=v_{X}+v_{Y}-v_{XY}$, i.e.,
the mutual information $I(X;Y;U)$. Without loss of generality, we assume $H(X)\ge H(Y)$.
Note that the original Gray--Wyner region and the region of tension correspond to
the upper-left corner.}
\end{figure}

\begin{figure}[h]
\begin{centering}
\includegraphics[scale=0.6]{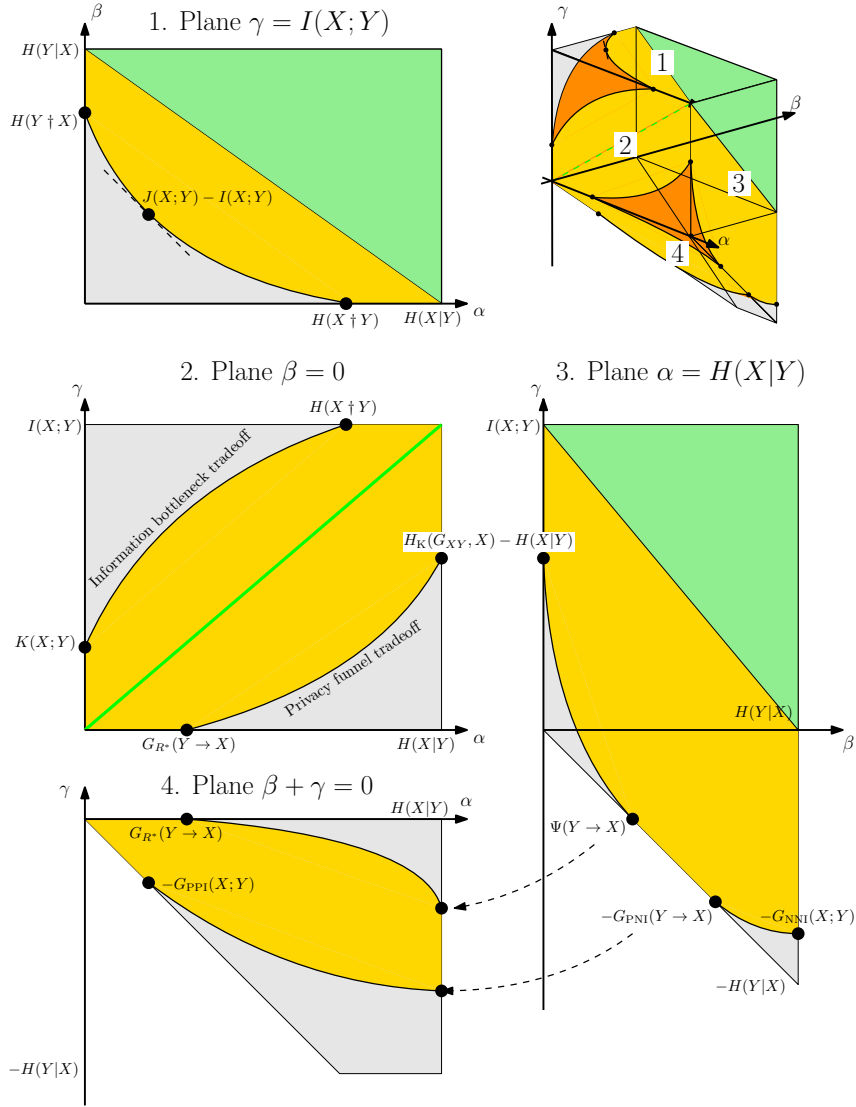}
\par\end{centering}
\caption{\label{fig:ixy_2d}Illustration of $\mathscr{I}_{XY}$ (yellow), $\mathscr{I}^{\mathrm{i}}_{XY}$
(green) and $\mathscr{I}^{\mathrm{o}}_{XY}$ (grey) restricted to different
planes. The axes are $\alpha=I(X;U|Y)=v_{XY}-v_{Y}$, $\beta=I(Y;U|X)=v_{XY}-v_{X}$
and $\gamma=v_{X}+v_{Y}-v_{XY}$. We assume $H(X)\ge H(Y)$.}

\end{figure}

Figures~\ref{fig:ixy_3d},~\ref{fig:ixy_2d} illustrate the mutual information region $\mathscr{I}_{XY}$ and its extreme points, and Table~\ref{table:extremepts} lists the extreme points and their corresponding optimization problems and support function representations. 

We first consider the extreme points of $\mathscr{I}_{XY}$ that correspond to previously known quantities. 
\smallskip

\noindent {\bf Wyner's common information}~\cite{Wyner1975a}
\[
J(X;Y)=\min_{X-U-Y}I(X,Y;U)
\]
can be expressed as
\begin{align*}
J(X;Y) & =\min\left\{ v_{XY}:\,v\in\mathscr{I}_{XY},\,v_{X}+v_{Y}-v_{XY}=I(X;Y)\right\} \\
 & =\min\left\{ R_{0}:\,R_{0}^{4}\in\mathscr{R},\,R_{0}+R_{1}+R_{2}=H(X,Y)\right\} \\
 & =-\psi'_{\mathscr{I}_{XY}}(1,1,-1;\,0,0,-1).
\end{align*}

\noindent {\bf G{\'a}cs-K{\"o}rner common information}~\cite{Gacs--Korner1973,witsenhausen1975sequences}
\[
K(X;Y)  =\max_{U:\,H(U|X)=H(U|Y)=0}H(U) =\max_{U:\,X-Y-U,\,U-X-Y}I(X,Y;U)
\]
can be expressed as
\begin{align*}
K(X;Y) & =\max\left\{ v_{XY}:\,v\in\mathscr{I}_{XY},\,v_{X}=v_{Y}=v_{XY}\right\} \\
 & =\max\left\{ R_{0}:\,R_{0}^{4}\in\mathscr{R},\,R_{0}+R_{1}=H(X),\,R_{0}+R_{2}=H(Y)\right\} \\
 & =\psi'_{\mathscr{I}_{XY}}(1,1,-2;\,0,0,1).
\end{align*}

\noindent {\bf K{\"o}rner graph entropy}~\cite{korner1973coding,alon1996}. Let
 $G_{XY}$ be a graph with a set of vertices $\mathcal{X}$ and edges between confusable
symbols upon observing $Y$, i.e., there is an edge $(x_{1},x_{2})$
if $p(x_{1},y),p(x_{2},y)>0$ for some $y$. The K{\"o}rner graph entropy 
\[
H_{\mathrm{K}}(G_{XY},X)  =\min_{U:\,U-X-Y,\,H(X|Y,U)=0}I(X;U)
\]
can be expressed as
\begin{align*}
H_{\mathrm{K}}(G_{XY},X) & =\min\left\{ v_{X}:\,v\in\mathscr{I}_{XY},\,v_{X}=v_{XY},\,v_{XY}-v_{Y}=H(X|Y)\right\} \\
 & =\min\left\{ R_{0}:\,R_{0}^{4}\in\mathscr{R},\,R_{0}+R_{1}=H(X),\,R_{3}=0\right\} \\
 & =-\psi'_{\mathscr{I}_{XY}}(1,-1,0;\,-1,0,0).
\end{align*}
In the Gray--Wyner system with causal complementary side information,
$H_{\mathrm{K}}(G_{XY},X)$ corresponds to the setting with only decoders
1, 3 and $M_{3}=\emptyset$, and we restrict the sum rate $R_{0}+R_{1}=H(X)$.
This is in line with the lossless source coding setting with causal
side information~\cite{Weissman--El-Gamal2006}, where the optimal
rate is also given by $H_{\mathrm{K}}(G_{XY},X)$. An intuitive reason
of this equality is that $R_{0}+R_{1}=H(X)$ and the recovery requirement
of decoder 1 forces $M_{0}$ and $M_{1}$ to
contain negligible information outside $X^{n}$, hence the setting
is similar to the case in which the encoder has access only to $X^{n}$. This corresponds to lossless source coding with
causal side information setting.
\smallskip

\noindent {\bf Necessary conditional entropy}~\cite{cuff2010coordination} (also
see $H(Y\searrow X|X)$ in~\cite{wolf2008monotones}, $G(Y\to X)$
in~\cite{kamath2010}, private information in~\cite{asoodeh2014privacy}
and~\cite{banerjee2015synergy})
\[
H(Y\dagger X)  =\min_{U:\,H(U|Y)=0,\,X-U-Y}H(U|X)
  =\min_{U:\,X-Y-U,\,X-U-Y}I(Y;U)-I(X;Y)
\]
can be expressed as
\begin{align*}
H(Y\dagger X) & =\min\left\{ v_{XY}:\,v\in\mathscr{I}_{XY},\,v_{Y}=v_{XY},\,v_{X}=I(X;Y)\right\} -I(X;Y)\\
 & =\min\left\{ R_{0}:\,R_{0}^{4}\in\mathscr{R},\,R_{0}+R_{2}=H(Y),\,R_{1}=H(X|Y)\right\} \\
 & =-\psi'_{\mathscr{I}_{XY}}(1,2,-2;\,1,0,-1).
\end{align*}

\noindent {\bf Information bottleneck}~\cite{tishby2000information}
\[
G_{\mathrm{IB}}(t,\,X\to Y)=\min_{U:\,X-Y-U,\,I(X;U)\ge t}I(Y;U)
\]
can be expressed as
\begin{align*}
G_{\mathrm{IB}}(t,\,X\to Y) & =\min\left\{ v_{Y}:\,v\in\mathscr{I}_{XY},\,v_{Y}=v_{XY},\,v_{X}\ge t\right\} \\
 & =\min\left\{ R_{0}:\,R_{0}^{4}\in\mathscr{R},\,R_{0}+R_{2}=H(Y),\,R_{1}\le H(X)-t\right\} .
\end{align*}
Note that the same tradeoff also appears in common randomness extraction
on a 2-DMS with one-way communication~\cite{ahlswede1998cr}, lossless
source coding with a helper~\cite{Wyner1973b,Ahlswede--Korner1975,Wyner1975b},
and a quantity studied by Witsenhausen and Wyner~\cite{witsenhausen1975conditional}.
It is shown in~\cite{Anantharam2013maxcor} that its slope is given
by the chordal slope of the hypercontractivity of Markov operator~\cite{ahlswede1976spreading}
\begin{align*}
s^{*}(Y\to X) & =\sup_{U:\,X-Y-U}\frac{I(X;U)}{I(Y;U)}\\
 & =\sup\left\{ v_{X}/v_{Y}:\,v\in\mathscr{I}_{XY},\,v_{Y}=v_{XY}\right\} .
\end{align*}

\noindent {\bf Privacy funnel}~\cite{makhdoumi2014funnel} (also see the rate-privacy
function defined in~\cite{asoodeh2014privacy})
\[
G_{\mathrm{PF}}(t,\,X\to Y)=\min_{U:\,X-Y-U,\,I(Y;U)\ge t}I(X;U)
\]
can be expressed as
\begin{align*}
G_{\mathrm{PF}}(t,\,X\to Y) & =\min\left\{ v_{X}:\,v\in\mathscr{I}_{XY},\,v_{Y}=v_{XY},\,v_{Y}\ge t\right\} \\
 & =\min\left\{ R_{0}+R_{4}-H(Y|X):\,R_{0}^{4}\in\mathscr{R},\,R_{0}+R_{2}=H(Y),\,R_{0}\ge t\right\} .
\end{align*}
In particular, the maximum $R$ for perfect privacy (written as $g_{0}(X;Y)$
in \cite{asoodeh2014privacy}, also see \cite{calmon2015perfect})
is
\begin{align*}
G_{R^{*}}(X\to Y) & =\max\left\{ t\ge0:\,G_{\mathrm{PF}}(t,\,X\to Y)=0\right\} \\
 & =\max\left\{ v_{Y}:\,v\in\mathscr{I}_{XY},\,v_{Y}=v_{XY},\,v_{X}=0\right\} \\
 & =\max\left\{ R_{0}:\,R_{0}^{4}\in\mathscr{R},\,R_{0}+R_{2}=H(Y),\,R_{0}+R_{4}=H(Y|X)\right\} \\
 & =\psi'_{\mathscr{I}_{XY}}(-1,1,-1;\,0,1,0).
\end{align*}
The optimal privacy-utility coefficient~\cite{calmon2015perfect} is
\begin{align*}v^{*}(X\to Y) & =\inf_{U:\,X-Y-U}\frac{I(X;U)}{I(Y;U)}\\
 & =\inf\left\{ v_{X}/v_{Y}:\,v\in\mathscr{I}_{XY},\,v_{Y}=v_{XY}\right\} .
\end{align*}

\noindent {\bf Excess functional information}~\cite{sfrl_arxiv} 
\[
\Psi(X\to Y)=\min_{U:\,U\perp\!\!\!\perp X}H(Y|U)-I(X;Y)
\]
is closely related to one-shot channel simulation~\cite{harsha2010communication} and lossy source
coding, and can be expressed as
\begin{align*}
\Psi(X\to Y) & =H(Y|X)-\max\left\{ v_{Y}:\,v\in\mathscr{I}_{XY},\,v_{X}=0\right\} \\
 & =\min\left\{ R_{2}:\,R_{0}^{4}\in\mathscr{R},\,R_{0}+R_{4}=H(Y|X)\right\} -I(X;Y)\\
 & =\min\left\{ R_{2}:\,R_{0}^{4}\in\mathscr{R},\,R_{4}=0,\,R_{0}=H(Y|X)\right\} -I(X;Y)\\
 & =-\psi'_{\mathscr{I}_{XY}}(-2,0,1;\,0,1,-1).
\end{align*}
In the EGW system,
$\Psi(X\to Y)$ corresponds to the setting with only decoders 2, 4
and $M_{4}=\emptyset$ (since it is better to allocate the rate to
$R_{0}$ instead of $R_{4}$), and we restrict $R_{0}=H(Y|X)$. The
value of $\Psi(X\to Y)+I(X;Y)$ is the rate of the additional information
$M_{2}$ that decoder 2 needs, in order to compensate the lack of
side information compared to decoder 4.
\smallskip

\noindent {\bf Minimum communication rate for distributed channel synthesis with common
randomness rate $t$}~\cite{cuff2013distributed,bennet2014reverse} 
\[
C(t,X\to Y)=\min_{U:\,X-U-Y}\max\left\{ I(X;U),\,I(X,Y;U)-t\right\} 
\]
can be expressed as
\begin{align*}
C(t,X\to Y) & =\min\left\{ \max\{v_{X},\,v_{XY}-t\}:\,v\in\mathscr{I}_{XY},\,v_{X}+v_{Y}-v_{XY}=I(X;Y)\right\} \\
 & =\min\left\{ \max\{H(X)-R_{1},\,R_{0}-t\}:\,R_{0}^{4}\in\mathscr{R},\,R_{0}+R_{1}+R_{2}=H(X,Y)\right\} .
\end{align*}

\subsection{New information theoretic quantities}

We now present three new quantities which arise as extreme points of $\mathscr{I}_{XY}$. These extreme points
concern the case in which decoders 3 and 4 are active in the EGW system. Note that they
are all maximizations of the interaction information $I(X;Y|U)-I(X;Y)$
under various constraints. They can be considered as distances from
extreme dependency, in the sense that they are equal to zero only under certain
conditions of extreme dependency.
\smallskip

\noindent {\bf Maximal interaction information} is defined as
\[
G_{\mathrm{NNI}}(X;\,Y)=\max_{p_{U|XY}}I(X;Y|U)-I(X;Y).
\]
It can be shown that
\begin{align*}
G_{\mathrm{NNI}}(X;\,Y) & =H(X|Y)+H(Y|X)-\min_{U:\,H(Y|X,U)=H(X|Y,U)=0}I(X,Y;U)\\
& = \max\left\{ v_{XY}-v_X-v_Y:\,v\in\mathscr{I}_{XY} \right\} \\
 & =H(X|Y)+H(Y|X)-\min\left\{ R_{0}+R_{3}+R_{4}:\,R_{0}^{4}\in\mathscr{R}\right\} \\
 & =H(X|Y)+H(Y|X)-\min\left\{ R_{0}:\,R_{0}^{4}\in\mathscr{R},\,R_{3}=R_{4}=0\right\} \\
 & = \psi_{\mathscr{I}_{XY}}(-1,-1,1).
\end{align*}
The maximal interaction information concerns the sum-rate of the EGW system with only decoders
3,4. Note that it is always better to allocate the rates $R_{3},R_{4}$
to $R_{0}$ instead, hence we can assume $R_{3}=R_{4}=0$ (which corresponds
to $H(Y|X,U)=H(X|Y,U)=0$). The quantity $H(X|Y)+H(Y|X)-G_{\mathrm{NNI}}(X;\,Y)$
is the maximum rate  in the lossless causal version of the complementary delivery setup~\cite{Kimura--Uyematsu2006}.
\smallskip

\noindent {\bf Asymmetric private interaction information} is defined as
\[
G_{\mathrm{PNI}}(X\to Y)=\max_{U:\,U\perp\!\!\!\perp X}I(X;Y|U)-I(X;Y).
\]
It can be shown that
\begin{align*}
G_{\mathrm{PNI}}(X\to Y) & =H(Y|X)-\min_{U:\,U\perp\!\!\!\perp X,\,H(Y|X,U)=0}I(Y;U)\\
 & =H(Y|X)-\min\left\{ v_{Y}:\,v\in\mathscr{I}_{XY},\,v_{X}=0,\,v_{XY}=H(Y|X)\right\} \\
 & =H(X|Y)-\min\left\{ R_{3}:\,R_{0}^{4}\in\mathscr{R},\,R_{0}+R_{4}=H(Y|X)\right\} \\
 & =H(X|Y)-\min\left\{ R_{3}:\,R_{0}^{4}\in\mathscr{R},\,R_{4}=0,\,R_{0}=H(Y|X)\right\} \\
 & =\psi'_{\mathscr{I}_{XY}}(-1,0,0;\,0,-1,1).
\end{align*}

The asymmetric private interaction information is the opposite of
excess functional information defined in~\cite{sfrl_arxiv} in which $I(Y;U)$ is maximized
instead. Another operational meaning of $G_{\mathrm{PNI}}$ is the
generation of random variables with a privacy constraint. Suppose
Alice observes $X$ and wants to generate $Y\sim p_{Y|X}(\cdot|X)$.
However, she does not have any private randomness and can only access
public randomness $W$, which
is also available to Eve. Her goal is to generate $Y$ as a function
of $X$ and $W$, while minimizing Eve's knowledge on $Y$ measured
by $I(Y;W)$. The minimum $I(Y;W)$ is $H(Y|X)-G_{\mathrm{PNI}}(X\to Y)$.
\smallskip

\noindent {\bf Symmetric private interaction information} is defined as
\[
G_{\mathrm{PPI}}(X;\,Y)=\max_{U:\,U\perp\!\!\!\perp X,\,U\perp\!\!\!\perp Y}I(X;Y|U)-I(X;Y).
\]
It can be shown that
\begin{align*}
G_{\mathrm{PPI}}(X;\,Y) & =\max_{U:\,U\perp\!\!\!\perp X,\,U\perp\!\!\!\perp Y}I(X,Y;U)\\
 & =\max\left\{ v_{XY}:\,v\in\mathscr{I}_{XY},\,v_{X}=v_{Y}=0\right\} \\
 & =\max\left\{ R_{0}:\,R_{0}^{4}\in\mathscr{R},\,R_{0}+R_{3}=H(X|Y),\,R_{0}+R_{4}=H(Y|X)\right\} \\
 & =\psi'_{\mathscr{I}_{XY}}(-1,-1,0;\,0,0,1).
\end{align*}

Intuitively, $G_{\mathrm{PPI}}$ captures the maximum amount of information
one can disclose about $(X,Y$), such that an eavesdropper who only
has one of $X$ or $Y$ would know nothing about the disclosed information.
Another operational meaning of $G_{\mathrm{PNI}}$ is the generation
of random variables with a privacy constraint (similar to that for
$G_{\mathrm{PNI}}$). Suppose Alice observes $X$ and wants to generate
$Y\sim p_{Y|X}(\cdot|X)$. She has access to public randomness $W$, which is also available to
Eve. She also has access to private randomness. Her goal is to generate
$Y$ using $X$, $W$ and her private randomness such that Eve has
no knowledge on $Y$ (i.e., $I(Y;W)=0$), while minimizing the amount
of private randomness used measured by $H(Y|X,W)$ (note that if Alice
can flip fair coins for the private randomness, then by Knuth-Yao algorithm~\cite{knuth1976complexity} the expected
number of flips is bounded by $H(Y|X,W)+2$ ).
The minimum $H(Y|X,W)$ is $H(Y|X)-G_{\mathrm{PPI}}(X;Y)$.
\medskip

We now list several properties of $G_{\mathrm{NNI}}$, $G_{\mathrm{PNI}}$
and $G_{\mathrm{PPI}}$. 
\begin{prop}
\label{prop:gpni}$G_{\mathrm{NNI}}$, $G_{\mathrm{PNI}}$ and $G_{\mathrm{PPI}}$
satisfies
\begin{enumerate}
\item Bounds.
\[
0\le G_{\mathrm{PPI}}(X;\,Y)\le G_{\mathrm{PNI}}(X\to Y)\le G_{\mathrm{NNI}}(X;\,Y)\le\min\left\{ H(X|Y),\,H(Y|X)\right\} .
\]
\item Conditions for zero.
\begin{itemize}
\item $G_{\mathrm{NNI}}(X;\,Y)=0$ if and only if the characteristic bipartite
graph of $X,Y$ (i.e. vertices $\mathcal{X}\cup\mathcal{Y}$ with
edge $(x,y)$ if $p(x,y)>0$) does not contain paths of length 3,
or equivalently, $p(x|y)=1$ or $p(y|x)=1$ for all $x,y$ such that
$p(x,y)>0$.
\item $G_{\mathrm{PNI}}(X\to Y)=0$ if and only if $G_{\mathrm{NNI}}(X;\,Y)=0$.
\item $G_{\mathrm{PPI}}(X;\,Y)=0$ if and only if the characteristic bipartite
graph of $X,Y$ does not contain cycles.
\end{itemize}
\item Condition for maximum. If $H(X)=H(Y)$, then the following statements are equivalent:
\begin{itemize}
\item $G_{\mathrm{NNI}}(X;\,Y)=H(Y|X)$.
\item $G_{\mathrm{PNI}}(X\to Y)=H(Y|X)$.
\item $G_{\mathrm{PPI}}(X;\,Y)=H(Y|X)$.
\item $p(x)=p(y)$ for all $x,y$ such that $p(x,y)>0$.
\end{itemize}
\item Lower bound for independent $X,Y$. If $X\perp\!\!\!\perp Y$,
\[
G_{\mathrm{PPI}}(X;\,Y)\ge\E\left[-\log \max\{p(X),\,p(Y)\}\right]-1.
\]
\item Superadditivity. If $(X_{1},Y_{1})$ is independent of $(X_{2},Y_{2})$,
then
\[
G_{\mathrm{NNI}}(X_{1},X_{2};\,Y_{1},Y_{2})\ge G_{\mathrm{NNI}}(X_{1};\,Y_{1})+G_{\mathrm{NNI}}(X_{2};\,Y_{2}).
\]
Similar for $G_{\mathrm{PNI}}$ and $G_{\mathrm{PPI}}$.
\end{enumerate}
\end{prop}
The proof of this proposition is given in Appendix~\ref{subsec:pf_iireg_prop}.

\begin{table}
\begin{center}
\begin{tabular}{|c|c|c|c|}
\hline 
$\!\!\!\begin{array}{c}
\text{Active}\\
\text{decoders}\\
\text{in EGW}
\end{array}\!\!\!$ & Information quantity & Objective and constraints in EGW & $\begin{array}{c}
\text{Support fcn. rep.}\\
(\psi=\psi_{\mathscr{I}_{XY}})
\end{array}$\tabularnewline
\hline 
\hline 
\multirow{5}{*}{$\!\!\!\begin{array}{c}
\text{1, 2}\end{array}\!\!\!$} & Wyner's CI~\cite{Wyner1975a} & $\min R_{0}:\,R_{0}+R_{1}+R_{2}=H(X,Y)$ & $-\psi'(1,1,-1;\,0,0,-1)$\tabularnewline
\cline{2-4} 
 & G{\'a}cs-K{\"o}rner CI~\cite{Gacs--Korner1973,witsenhausen1975sequences} & $\max R_{0}:\,R_{0}+R_{1}=H(X),\,R_{0}+R_{2}=H(Y)$ & $\psi'(1,1,-2;\,0,0,1)$\tabularnewline
\cline{2-4} 
 & Necessary conditional entropy~\cite{cuff2010coordination,wolf2008monotones} & $\min R_{0}:\,R_{0}+R_{2}=H(Y),\,R_{1}=H(X|Y)$ & $-\psi'(1,2,-2;\,1,0,-1)$\tabularnewline
\cline{2-4} 
 & Info. bottleneck~\cite{tishby2000information} & $\min R_{0}:\,R_{0}+R_{2}=H(Y),\,R_{1}\le H(X)-t$ & none\tabularnewline
\cline{2-4} 
 & Comm. rate for channel synthesis~\cite{cuff2013distributed,bennet2014reverse} & $ \min \max\{H(X)\!-\! R_{1},R_{0}\!-\! t\}:R_{0}\!+\! R_{1}\!+\! R_{2}\!=\! H(X,Y)$ & none\tabularnewline
\hline 
\hline 
\multirow{4}{*}{$\!\!\!\begin{array}{c}
\text{1, 3}\\
\text{or 2, 4}
\end{array}\!\!\!$} & K{\"o}rner graph entropy~\cite{korner1973coding} & $\min R_{0}:\,R_{0}+R_{1}=H(X),\,R_{3}=0$ & $-\psi'(1,-1,0;\,-1,0,0)$\tabularnewline
\cline{2-4} 
 & Excess functional info.~\cite{sfrl_arxiv} & $\min R_{2}-I(X;Y):\,R_{4}=0,\,R_{0}=H(Y|X)$ & $-\psi'(-2,0,1;\,0,1,-1)$\tabularnewline
\cline{2-4} 
 & Max. rate for perfect privacy~\cite{makhdoumi2014funnel,asoodeh2014privacy} & $\max R_{0}:\,R_{0}+R_{2}=H(Y),\,R_{0}+R_{4}=H(Y|X)$ & $\psi'(-1,1,-1;\,0,1,0)$\tabularnewline
\cline{2-4} 
 & Privacy funnel~\cite{makhdoumi2014funnel} & $\min R_{0}+R_{4}-H(Y|X):\,R_{0}+R_{2}=H(Y),\,R_{0}\ge t$ & none\tabularnewline
\hline 
\hline 
\multirow{3}{*}{$\!\!\!\begin{array}{c}
\text{3, 4}\end{array}\!\!\!$} & Maximal interaction info. & $\max H(X|Y)+H(Y|X)-R_{0}:\,R_{3}=R_{4}=0$ & $\psi(-1,-1,1)$\tabularnewline
\cline{2-4} 
 & Asymm. private interaction info. & $\max H(X|Y)-R_{3}:\,R_{4}=0,\,R_{0}=H(Y|X)$ & $\psi'(-1,0,0;\,0,-1,1)$\tabularnewline
\cline{2-4} 
 & Symm. private interaction info. & $\max R_{0}:\,R_{0}+R_{3}=H(X|Y),\,R_{0}+R_{4}=H(Y|X)$ & $\psi'(-1,-1,0;\,0,0,1)$\tabularnewline
\hline 
\end{tabular}
\smallskip

\caption{Extreme points of $\mathscr{I}_{XY}$ and the corresponding extreme points
in the EGW, and their support function representations. }
\label{table:extremepts}
\end{center}
\end{table}

\section{Extended Gray--Wyner system with Noncausal Complementary Side Information\label{sec:asymp}}

In this section we establish the rate region $\Rr'$ for the EGW system
with complementary noncausal side information at decoders 3 and 4
(noncausal EGW), that is, decoder 3 recovers $X^{n}$ from $(M_{0},M_{3},Y^{n})$
and decoder 4 similarly recovers $Y^{n}$ from $(M_{0},M_{4},X^{n})$.
We show that $\Rr'$ can be expressed in terms of the Gray-Wyner
region $\mathscr{R}_{\mathrm{GW}}$, hence it contains fewer interesting
extreme points compared to $\Rr$. This is the reason we emphasized the
causal side information in this paper.
We further show that $\Rr'$ is related to the \emph{asymptotic mutual
information region} defined as 
\[
\mathscr{I}_{XY}^{\infty}=\bigcup_{n=1}^{\infty}\frac{1}{n}\mathscr{I}_{X^{n},Y^{n}},
\]
where $(X^{n},Y^{n})$ is i.i.d. with $(X_{1},Y_{1})\sim p_{XY}$.
Note that $\mathscr{I}_{XY}^{\infty}$ may not be closed (unlike $\mathscr{I}_{XY}$
which is always closed).


The following gives the rate region for the noncausal EGW. 
\begin{thm}
\label{thm:gwn}The optimal rate region $\Rr'$ for the extended Gray--Wyner
system with noncausal complementary side information is the set of
rate tuples $(R_{0},R_{1},R_{2},R_{3},R_{4})$ such that 
\begin{align*}
R_{0} & \ge I(X,Y;U),\\
R_{1} & \ge H(X|U),\\
R_{2} & \ge H(Y|U),\\
R_{3} & \ge H(X|U)-H(Y),\\
R_{4} & \ge H(Y|U)-H(X),\\
R_{0}+R_{3} & \ge H(X|Y),\\
R_{0}+R_{4} & \ge H(Y|X),\\
R_{2}+R_{3} & \ge H(X|U),\\
R_{1}+R_{4} & \ge H(Y|U),\\
R_{0}+R_{2}+R_{3} & \ge H(X,Y),\\
R_{0}+R_{1}+R_{4} & \ge H(X,Y)
\end{align*}
for some $p_{U|XY}$, where $|\mathcal{U}|\le|\mathcal{X}|\cdot|\mathcal{Y}|+2$. 
\end{thm}
The proof is given in Appendix~\ref{subsec:pf_gwn}.
Then we characterize the closure of $\mathscr{I}_{XY}^{\infty}$.
We show that $\mathrm{cl}(\mathscr{I}_{XY}^{\infty})$, $\Rr'$ and
the the Gray--Wyner region $\Rr_{\mathrm{GW}}$ can be expressed
in terms of each other.
\begin{prop}
\label{prop:cl_ixy}The closure of $\mathscr{I}_{XY}^{\infty}$, the
rate region $\Rr'$ for the noncausal EGW and the Gray--Wyner
region $\mathscr{R}_{\mathrm{GW}}$ satisfy:
\begin{enumerate}
\item Characterization of $\mathrm{cl}(\mathscr{I}_{XY}^{\infty})$.
\begin{align*}
\mathrm{cl}(\mathscr{I}_{XY}^{\infty}) & =\left(\mathscr{I}_{XY}+(-\infty,0]\times(-\infty,0]\times[0,\infty)\right)\cap\mathscr{I}^{\mathrm{o}}_{XY}\\
&=\left(\mathscr{I}_{XY}+ \{(t, t, t) :\, t \le 0 \} \right)\cap \big( [0,\infty) \times [0,\infty) \times \mathbb{R} \big).
\end{align*}
\item Equivalence between $\mathrm{cl}(\mathscr{I}_{XY}^{\infty})$ and
$\Rr'$.
\begin{align*}
\Rr'=\bigcup_{v\in\mathrm{cl}(\mathscr{I}_{XY}^{\infty})} & \big\{\big(v_{XY},\,H(X)-v_{X},\,H(Y)-v_{Y},\,H(X|Y)-v_{XY}+v_{Y},\,H(Y|X)-v_{XY}+v_{X}\big)\big\}+[0,\infty)^{5},
\end{align*}
and
\begin{align*}
\mathrm{cl}(\mathscr{I}_{XY}^{\infty})=\left\{ v\in\mathbb{R}^{3}:\,\big(v_{XY},\,H(X)-v_{X},\,H(Y)-v_{Y},\,H(X|Y)-v_{XY}+v_{Y},\,H(Y|X)-v_{XY}+v_{X}\big)\in\Rr'\right\} .
\end{align*}
\item Equivalence between $\mathrm{cl}(\mathscr{I}_{XY}^{\infty})$ and
$\mathscr{R}_{\mathrm{GW}}$.
\begin{align*}
\mathscr{R}_{\mathrm{GW}} & =\bigcup_{v\in\mathrm{cl}(\mathscr{I}_{XY}^{\infty})}\big\{\big(v_{XY},\,H(X)-v_{X},\,H(Y)-v_{Y}\big)\big\}+[0,\infty)^{3},
\end{align*}
and
\[
\mathrm{cl}(\mathscr{I}_{XY}^{\infty})=\left\{ v\in\mathscr{I}^{\mathrm{o}}_{XY}:\,\big(v_{XY},\,H(X)-v_{X},\,H(Y)-v_{Y}\big)\in\mathscr{R}_{\mathrm{GW}}\right\} .
\]
\end{enumerate}
\end{prop}
The proof is given in Appendix~\ref{subsec:pf_cl_ixy}. Note that
Proposition~\ref{prop:cl_ixy} does not characterize $\mathscr{I}_{XY}^{\infty}$
completely since it does not specify which boundary points are in
$\mathscr{I}_{XY}^{\infty}$. 

\appendix

\subsection{Proof of the converse of Theorem \ref{thm:gwc}\label{subsec:pf_gwc}}
To prove the converse, let $U_{i}=(M_{0},X^{i-1},Y^{i-1})$. Consider{\allowdisplaybreaks
\begin{align}
nR_{0} & \ge I(X^{n},Y^{n};M_{0}) \nonumber\\
 & =\sum_{i=1}^{n}I(X_{i},Y_{i};M_{0}|X^{i-1},Y^{i-1})\nonumber\\
 & =\sum_{i=1}^{n}I(X_{i},Y_{i};M_{0},X^{i-1},Y^{i-1})\nonumber\\
 & =\sum_{i=1}^{n}I(X_{i},Y_{i};U_{i}),\nonumber\\
nR_{1} & \ge H(M_{1}|M_{0})\nonumber \\
 & \ge I(X^{n};M_{1}|M_{0})\nonumber \\
 & =H(X^{n}|M_{0})-H(X^{n}|M_{0},M_{1})\nonumber \\
 & =\sum_{i=1}^{n}H(X_{i}|M_{0},X^{i-1})-H(X^{n}|M_{0},M_{1})\nonumber \\
 & \ge\sum_{i=1}^{n}H(X_{i}|M_{0},X^{i-1},Y^{i-1})-H(X^{n}|M_{0},M_{1})\nonumber \\
 & \ge\sum_{i=1}^{n}H(X_{i}|M_{0},X^{i-1},Y^{i-1})-\log\left|\mathcal{X}\right|\sum_{i=1}^{n}\P\left\{ X_{i}\neq\hat{X}_{1,i}\right\} -1\label{eq:gwc_symerr}\\
 & \ge\sum_{i=1}^{n}H(X_{i}|U_{i})-o(n),\nonumber 
\end{align}
where the last inequality follows by Fano's inequality. Similarly $nR_{2}\ge\sum_{i}H(Y_{i}|U_{i})-o(n)$. Next, consider
\begin{align}
nR_{3} & \ge H(M_{3}|M_{0})\nonumber \\
 & \ge I(X^{n},Y^{n};M_{3}|M_{0})\nonumber \\
 & =\sum_{i=1}^{n}I(X_{i},Y_{i};M_{3}|M_{0},X^{i-1},Y^{i-1})\nonumber \\
 & \ge\sum_{i=1}^{n}I(X_{i};M_{3}|M_{0},X^{i-1},Y^{i})\nonumber \\
 & =\sum_{i=1}^{n}\left(H(X_{i}|M_{0},X^{i-1},Y^{i})-H(X_{i}|M_{0},M_{3},X^{i-1},Y^{i})\right)\nonumber \\
 & =\sum_{i=1}^{n}H(X_{i}|Y_{i},U_{i})-\sum_{i=1}^{n}H(X_{i}|M_{0},M_{3},X^{i-1},Y^{i})\nonumber \\
 & \ge\sum_{i=1}^{n}H(X_{i}|Y_{i},U_{i})-\log\left|\mathcal{X}\right|\sum_{i=1}^{n}\P\left\{ X_{i}\neq\hat{X}_{3,i}\right\} -1\label{eq:gwc_symerr2}\\
 & =\sum_{i=1}^{n}H(X_{i}|Y_{i},U_{i})-o(n),\nonumber 
\end{align}}
where the last inequality follows by Fano's inequality since $\hat{X}_{3,i}$
is a function of $M_{0},M_{3},Y^{i}$. Similarly $nR_{4}\ge\sum_{i}H(Y_{i}|X_{i},U_{i})-o(n)$.
Hence the point $(R_{0}+\epsilon,\ldots,R_{4}+\epsilon)$ is in the
convex hull of $\mathscr{R}$ for any $\epsilon>0$.
From \eqref{eq:gwc_ixy}, $\mathscr{R}$ is the increasing
hull of an affine transformation of $\mathscr{I}_{XY}$, and thus
is convex.

To prove the cardinality bound,
we apply Fenchel-Eggleston-Carath\'eodory
theorem~\cite{Eggleston1958,Rockafellar1970} on the $(|\Xc||\Yc|+2)$-dimensional
vectors with entries $H(X|U=u)$, $H(Y|U=u)$, $H(X,Y|U=u)$ and $p(x,y|u)$ for $u\in \{1,\ldots,|\mathcal{U}|\}$,  $(x,y)\in \{1,\ldots,|\Xc|\} \times \{1,\ldots,|\Yc|\} \backslash (|\Xc|,|\Yc|)$; see~\cite{Ahlswede--Korner1975,Wyner--Ziv1976}.

\subsection{Proof of Proposition \ref{prop:ixy_prop}\label{subsec:pf_ixy_prop}}
\begin{enumerate}
\item To see that $\mathscr{I}_{XY}$ is convex, for any $U_{0},U_{1}$
and $\lambda\in[0,1]$, let $Q\sim\mathrm{Bern}(\lambda)$ be independent
of $X,Y,U_{0},U_{1}$, and let $U=(Q,U_{Q})$. Then $I(X;U)=(1-\lambda)I(X;U_{0})+\lambda I(X;U_{1})$
(similarly for the other two quantities). Compactness will be proved later.

\item The outer bound follows directly from the properties of entropy and mutual information.

\item For the inner bound, the first 4 points can be obtained by substituting $U= \emptyset,\, X,\, Y,\, (X,Y)$ respectively. For the last point, by the functional representation lemma~\cite[p. 626]{elgamal2011network}, let ${V \perp \!\!\! \perp X}$ such that $H(Y|X,V)=0$. Again by the functional representation lemma, let ${W \perp \!\!\! \perp (Y,V)}$ such that $H(X|Y,V,W)=0$. Let $U=(V,W)$, then $I(X,Y;U)-I(X;U)=H(Y|X)-H(Y|X,U)=H(Y|X)$, $I(X,Y;U)-I(Y;U)=H(X|Y)-H(X|Y,U)=H(X|Y)$, and 
\begin{align*}
I(X,Y;U) & = I(X,Y;V,W) \\
& = I(X,Y;V) + I(X,Y;W|V)\\
& = I(Y;V|X) + I(X;W|Y,V)\\
& \le H(Y|X) + H(X|Y).
\end{align*}
Hence there exists $t\le H(Y|X) + H(X|Y)$ such that $(t-H(Y|X),\, t-H(X|Y),\,t) \in \mathscr{I}_{XY}$ (by substituting $t=I(X,Y;U)$). Taking convex combination of this point and $(H(X),H(Y),H(X,Y)) \in  \mathscr{I}_{XY}$, we have $(H(X|Y),H(Y|X),H(X|Y)+H(Y|X)) \in  \mathscr{I}_{XY}$.

The existence of $0\le\epsilon_{1}\le\log I(X;Y)+4$ such that
$(0,H(Y|X)-\epsilon_{1},H(Y|X)) \in \mathscr{I}_{XY}$
can be proved by substituting $\epsilon_{1} = \Psi(X \to Y)$ and invoking the strong functional representation lemma~\cite{sfrl_arxiv}.

\item The superadditivity property can be obtained from considering $U=(U_1,U_2)$, where $(I(X_i;U_i),\,I(Y_i;U_i),\,I(X_i,Y_i;U_i)) \in \mathscr{I}_{X_i,Y_i}$. 

\item The data processing property can be obtained from considering $U$ where $(I(X_1;U),\,I(Y_1;U),\,I(X_1,Y_1;U)) \in \mathscr{I}_{X_1,Y_1}$.

\item The cardinality bound can be proved using Fenchel-Eggleston-Carath\'eodory
theorem using the same arguments as in the converse proof of Theorem \ref{thm:gwc}. Compactness follows
from the fact that mutual information is a continuous function, and
the set of conditional pmfs $p_{U|XY}$ with $|\mathcal{U}|\le|\mathcal{X}|\cdot|\mathcal{Y}|+2$
is a compact set.

\item The relation to Gray--Wyner region and region of tension follows from the definitions of the regions.
\end{enumerate}

\subsection{Proof of Proposition~\ref{prop:gpni} \label{subsec:pf_iireg_prop}}

\begin{enumerate}

\item To prove the bound, note that $I(X;Y|U) \le H(X)$, hence $I(X;Y|U)-I(X;Y) \le H(X|Y)$, $G_{\mathrm{NNI}} \le H(X|Y)$.

\item We first prove that if there does not exist length 3 paths in the bipartite
graph, then $G_{\mathrm{NNI}}(X;\,Y)=G_{\mathrm{PNI}}(X;\,Y)=0$.
Let $Q$ achieves the G{\'a}cs-K{\"o}rner common information, i.e.,
$Q$ represents which connected component the edge $(X,Y)$ lies in.
If the bipartite graph does not contain length 3 paths, every connected
component is a star, i.e., for each $q$, either $H(X|Q=q)=0$ or
$H(Y|Q=q)=0$. Then $I(X;Y)=H(Q)+I(X;Y|Q)=H(Q)$, and $I(X;Y|U)=H(Q|U)+I(X;Y|Q,U)=H(Q|U)\le H(Q)$
for any $U$. Hence $G_{\mathrm{NNI}}(X;\,Y)=G_{\mathrm{PNI}}(X;\,Y)=0$.

We then prove that if there exist a length 3 path in the bipartite graph,
then $G_{\mathrm{NNI}}(X;\,Y)\ge G_{\mathrm{PNI}}(X;\,Y)>0$. Assume
$p(x_{1},y_{1}),p(x_{1},y_{2}),p(x_{2},y_{1})>0$. Let $U\in\{1,2\}$,
\[
p(u|x,y)=\begin{cases}
1/2+\epsilon/p(x_{1},y_{1}) & \text{if}\;(x,y,u)=(x_{1},y_{1},1)\\
1/2-\epsilon/p(x_{1},y_{1}) & \text{if}\;(x,y,u)=(x_{1},y_{1},2)\\
1/2-\epsilon/p(x_{1},y_{2}) & \text{if}\;(x,y,u)=(x_{1},y_{2},1)\\
1/2+\epsilon/p(x_{1},y_{2}) & \text{if}\;(x,y,u)=(x_{1},y_{2},2)\\
1/2 & \text{otherwise},
\end{cases}
\]
where $\epsilon>0$ is small enough such that the above is a valid
conditional pmf. One can verify that $U\perp\!\!\!\perp X$. Since
$p_{U|XY}(1|x_{1},y_{1})=1/2+\epsilon/p(x_{1},y_{1})\neq1/2=p_{U|XY}(1|x_{2},y_{1})$,
$X$ and $U$ are not conditionally independent given $Y$. Hence
$I(X;Y|U)-I(X;Y)=I(X;U|Y)>0$.

We then prove that if $G_{\mathrm{PPI}}(X;\,Y)>0$, then there exists
a cycle in the bipartite graph. Let $U$ satisfies $U\perp\!\!\!\perp X$,
$U\perp\!\!\!\perp Y$ and $I(X;Y|U)>I(X;Y)$. Since $U$ is not independent
of $X,Y$, there exists $x_{1},y_{1},u$ such that $p(x_{1},y_{1}|u)>p(x_{1},y_{1})$.
Since $\sum_{y'}p(x_{1},y'|u)=p(x_{1}|u)=p(x_{1})=\sum_{y'}p(x_{1},y')$,
there exists $y_{2}\neq y_{1}$ such that $p(x_{1},y_{2}|u)<p(x_{1},y_{2})$.
Since $\sum_{x'}p(x',y_{2}|u)=p(y_{2}|u)=p(y_{2})=\sum_{x'}p(x',y_{2})$,
there exists $x_{2}\neq x_{1}$ such that $p(x_{2},y_{2}|u)>p(x_{2},y_{2})$.
Continue this process until we return to a visited $x,y$ pair, i.e.,
$(x_{a},y_{a})=(x_{b},y_{b})$ for $a<b$. Then $y_{a},x_{a},y_{a+1},x_{a+1},\ldots,x_{b-1},y_{b}$
forms a cycle.

We then prove that if there exist a cycle in the bipartite graph, then
$G_{\mathrm{PPI}}(X;\,Y)>0$. Let $y_{1},x_{1},y_{2},x_{2},\ldots,x_{a},y_{a+1}=y_{1}$
be a cycle. Let $U\in\{1,2\}$, 
\[
p(u|x,y)=\begin{cases}
1/2+\epsilon/p(x_{i},y_{i}) & \text{if}\;(x,y,u)=(x_{i},y_{i},1)\\
1/2-\epsilon/p(x_{i},y_{i}) & \text{if}\;(x,y,u)=(x_{i},y_{i},2)\\
1/2-\epsilon/p(x_{i},y_{i+1}) & \text{if}\;(x,y,u)=(x_{i},y_{i+1},1)\\
1/2+\epsilon/p(x_{i},y_{i+1}) & \text{if}\;(x,y,u)=(x_{i},y_{i+1},2)\\
1/2 & \text{otherwise},
\end{cases}
\]
where $\epsilon>0$ is small enough such that the above is a valid
conditional pmf. One can verify that $U\perp\!\!\!\perp X$ and $U\perp\!\!\!\perp Y$.
Since $p_{U|XY}(1|x_{1},y_{1})>1/2>p_{U|XY}(1|x_{1},y_{2})$, $U$
is not independent of $X,Y$. Hence $I(X;Y|U)-I(X;Y)=I(X,Y;U)>0$.

\item We then prove that if $H(X)=H(Y)$ and $p(x)=p(y)$ for all $x,y$
such that $p(x,y)>0$, then $G_{\mathrm{PPI}}(X;\,Y)=H(Y|X)$. Let
$Q$ achieves the G{\'a}cs-K{\"o}rner common information, and let
$\mathcal{X}_{q}=\left\{ x:\,p(x|q)>0\right\} $, $\mathcal{Y}_{q}=\left\{ y:\,p(y|q)>0\right\} $,
then $X|\{Q=q\}\sim\mathrm{Unif}(\mathcal{X}_{q})$, $Y|\{Q=q\}\sim\mathrm{Unif}(\mathcal{Y}_{q})$
and $|\mathcal{X}_{q}|=|\mathcal{Y}_{q}|$ for all $q$. Applying
Birkhoff-von Neumann theorem on the submatrix of $p(x,y)$ with rows
$\mathcal{X}_{q}$ and columns $\mathcal{Y}_{q}$, there exists $U_{q}$
such that $p(x,y|q)=\sum_{u}p_{U_{q}}(u)p_{XY|U_{q}Q}(x,y|u,q)$,
$p_{X|U_{q}Q}(x|u,q)=p_{Y|U_{q}Q}(y|u,q)=1/|\mathcal{X}_{q}|$ and
$p_{XY|U_{q}Q}(x,y|u,q)\in\{0,\,1/|\mathcal{X}_{q}|\}$ for all $x,y,u$.
Let $U=\{U_{q}\}_{q\in\mathcal{Q}}$, where $U_{q}$ are assumed to
be independent across $q$. Then for any $x$ and $u=\{u_{q}\}$,
\begin{align*}
p(x|\{u_{q}\}) & =p(x,q|\{u_{q}\})\\
 & =p(q)p(x|u_{q},q)\\
 & =p(q)/|\mathcal{X}_{q}|\\
 & =p(x),
\end{align*}
where $q=q(x)$ since $H(Q|X)=0$. Hence $U\perp\!\!\!\perp X$. Similarly
$U\perp\!\!\!\perp Y$. Also since there is only one non-zero in $p_{XY|U_{q}Q}(x,y|u,q)$
for different $x$, we have $H(X|Y,U)=0$. Similarly $H(Y|X,U)=0$.
Hence $I(X;Y|U)-I(X;Y)=I(Y;U|X)-I(Y;U)=H(Y|X)$.

We then prove that if $H(X)=H(Y)$ and $G_{\mathrm{NNI}}(X;\,Y)=H(Y|X)$,
then $p(x)=p(y)$ for all $x,y$ such that $p(x,y)>0$. Let $U$ satisfies
$I(X;Y|U)=I(X;Y)+H(Y|X)=H(Y)$, then one can check that $U\perp\!\!\!\perp X$,
$U\perp\!\!\!\perp Y$, $H(X|Y,U)=0$ and $H(Y|X,U)=0$. For any $x,y$
such that $p(x,y)>0$, let $u$ such that $p(x,y,u)>0$, then 
\begin{align*}
p(x) & =p(x|u)\\
 & =p(x|u)p(y|x,u)\\
 & =p(y|u)p(x|y,u)\\
 & =p(y).
\end{align*}

\item We then prove the lower bound when $X,Y$ independent. Assume $X\perp\!\!\!\perp Y$.
Assume $\mathcal{X}=\{1,\ldots,|\mathcal{X}|\}$, $Y=\{1,\ldots,|\mathcal{Y}|\}$,
$X=F_{X}^{-1}(V)$, $Y=F_{Y}^{-1}(W)$, $V,W\sim\mathrm{Unif}[0,1]$
independent. Let $U=V+W\;\mathrm{mod}\;1$, then $U\perp\!\!\!\perp X$,
$U\perp\!\!\!\perp Y$.{\allowdisplaybreaks
\begin{align*}
H(Y|U,X) & =\sum_{x}p(x)\int_{0}^{1}H(Y|U=u,X=x)du\\
 & =\sum_{x}p(x)\int_{0}^{1}H(Y\,|\,W\in([u-F_{X}(x),\,u-F_{X}(x-1))\;\mathrm{mod}\;1))du\\
 & =\sum_{x}p(x)\int_{0}^{1}H(Y\,|\,W\in([u,\,u+p(x)]\;\mathrm{mod}\;1))du\\
 & =\sum_{x}p(x)\int_{0}^{1}\sum_{y}l\left(\P\left\{ Y=y\,|\,W\in([u,\,u+p(x)]\;\mathrm{mod}\;1)\right\} \right)du\\
 & =\sum_{x}p(x)\int_{0}^{1}\sum_{y}l\left(p(x)^{-1}\left|[F_{Y}(y-1),\,F_{Y}(y)]\cap([u,\,u+p(x)]\;\mathrm{mod}\;1)\right|\right)du\\
 & =\sum_{x}p(x)\sum_{y}\int_{0}^{1}l\left(p(x)^{-1}\left|[0,\,p(y)]\cap([u,\,u+p(x)]\;\mathrm{mod}\;1)\right|\right)du\\
 & =-H(X)+\sum_{x,y}\int_{0}^{1}l\left(\left|[0,\,p(y)]\cap([u,\,u+p(x)]\;\mathrm{mod}\;1)\right|\right)du
\end{align*}
where we write $A\;\mathrm{mod}\;1=\left\{ a\;\mathrm{mod}\;1:\,a\in A\right\} $
and $|A|$ for the Lebesgue measure for $A\subseteq\mathbb{R}$, $l(t)=-t\log t$.
Consider
\[
f(a,b)=\int_{0}^{1}l\left(\left|[0,\,b]\cap([u,\,u+a]\;\mathrm{mod}\;1)\right|\right)du.
\]
If $b\le a\le1$ and $a+b\le1$,
\begin{align*}
f(a,b) & =(a-b)l(b)+2\int_{0}^{b}l\left(u\right)du\\
 & \le(a-b)l(b)+2bl\left(b/2\right)\\
 & =al(b)+b^{2}\\
 & =ab\log\frac{1}{b}+b^{2}\\
 & \le ab\log\frac{1}{b}+ab.
\end{align*}
If $b\le a$ and $a+b>1$,
\begin{align*}
f(a,b) & =(a-b)l(b)+(a+b-1)l(a+b-1)+2\int_{b+a-1}^{b}l(u)du\\
 & \le(a-b)l(b)+(a+b-1)l(a+b-1)+2(1-a)l\left(b-\frac{1-a}{2}\right)\\
 & \le(a-b)l(b)+(1+b-a)l\left(\frac{b^{2}}{1+b-a}\right)\\
 & =(a-b)l(b)+b^{2}\log\frac{1+b-a}{b^{2}}\\
 & \le(a-b)l(b)+b^{2}\log\frac{2b}{b^{2}}\\
 & =al(b)+b^{2}\\
 & \le ab\log\frac{1}{b}+ab.
\end{align*}
Hence
\begin{align*}
I(X,Y;U) & =H(X,Y)-H(Y|U,X)-H(X|U)\\
 & =H(X,Y)-\sum_{x,y}f(p(x),p(y))\\
 & \ge H(X,Y)-\sum_{x,y}\left(p(x)p(y)\log\frac{1}{\min\{p(x),p(y)\}}+p(x)p(y)\right)\\
 & =\E\left[\log\frac{1}{\max\{p(X),p(Y)\}}\right]-1.
\end{align*}}

\item The superadditivity property follows from the superadditivity of mutual information region.

\end{enumerate}

\subsection{Proof of Theorem \ref{thm:gwn}\label{subsec:pf_gwn}}

We first prove the achievability. Without loss of generality assume
$H(X)\ge H(Y)$. Fix any point $v=(v_{X},v_{Y},v_{XY})\in\mathscr{I}_{XY}$.
Consider the region
\[
\mathscr{I}(v)=\left((-\infty,v_{X}]\times(-\infty,v_{Y}]\times[v_{XY},\infty)\right)\cap\mathscr{I}^{\mathrm{o}}_{XY}.
\]
It can be seen from Figure \ref{fig:ixy_3d} that $\mathscr{I}(v)$
is a subset of the convex hull of the following 9 points:
\begin{align*}
 & v,\\
 & p_{1}=(0,\,0,\,0),\\
 & p_{2}=(H(X),\,I(X;Y),\,H(X)),\\
 & p_{3}=(I(X;Y),\,H(Y),\,H(Y)),\\
 & p_{4}=(H(X),\,H(Y),\,H(X,Y)),\\
 & p_{5}=(H(X|Y),\,0,\,H(X|Y)),\\
 & p_{6}=(0,\,H(Y|X),\,H(Y|X)),\\
 & p_{7}=(0,\,0,\,H(Y|X)),\\
 & p_{8}=(H(X)-H(Y),\,0,\,H(X|Y)),
\end{align*}
i.e., $v$ together with the corner points of $\mathscr{I}^{\mathrm{o}}_{XY}$
except $(I(X;Y),\,I(X;Y),\,I(X;Y))$. We will prove that for any $w=(w_{X},w_{Y},w_{XY})\in\mathscr{I}(v)$,
the rate tuple $R(w)=(R_{0}(w),\ldots,R_{4}(w))$,
\begin{align*}
R_{0}(w) & =w_{XY}+\epsilon,\\
R_{1}(w) & =H(X)-w_{X}+\epsilon,\\
R_{2}(w) & =H(Y)-w_{Y}+\epsilon,\\
R_{3}(w) & =H(X|Y)-w_{XY}+w_{Y}+\epsilon,\\
R_{4}(w) & =H(Y|X)-w_{XY}+w_{X}+\epsilon
\end{align*}
is achievable in the extended Gray--Wyner system with noncausal complementary
side information for $\epsilon>0$. It suffices to prove the corner
points $R(v),R(p_{1}),\ldots,R(p_{8})$ are achievable.

$R(v)$ is achievable using the causal scheme in Theorem \ref{thm:gwc}.
To achieve $R(p_{1})$, $R(p_{2})$, $R(p_{3})$ and $R(p_{4})$,
apply the causal scheme in Theorem \ref{thm:gwc} on $U\leftarrow\emptyset$,
$U\leftarrow X$, $U\leftarrow Y$ and $U\leftarrow(X,Y)$, respectively.

To achieve $R(p_{5})$, applying the strong functional representation
lemma~\cite{sfrl_arxiv}, there exists $V_{n}\perp\!\!\!\perp Y^{n}$ such
that $H(X^{n}|Y^{n},V_{n})=0$ and $I(V_{n};Y^{n}|X^{n})\le\epsilon n/2$
for $n$ large enough. We then apply the causal scheme on $X\leftarrow X^{n}$,
$Y\leftarrow Y^{n}$ and $U\leftarrow V_{n}$. Similar for $R(p_{6})$.\medskip{}

We now prove the achievability of $R(p_{7})$. To generate the codebook,
randomly partition $\mathcal{T}_{\epsilon'}^{(n)}(X,Y)$ into bins
$\mathcal{B}_{0}(m_{0})$ of size $2^{n(H(X,Y)+\epsilon/2-R_{0})}$
for $m_{0}\in[1:\,2^{nR_{0}}]$. Further randomly partition the bin
$\mathcal{B}_{0}(m_{0})$ into $\mathcal{B}_{3}(m_{0},m_{3})$ of
size $2^{n(H(X,Y)+\epsilon/2-R_{0}-R_{3})}$ for $m_{3}\in[1:\,2^{nR_{3}}]$.

To encode $x^{n},y^{n}$, find $m_{0},m_{3}$ such that $(x^{n},y^{n})\in\mathcal{B}_{3}(m_{0},m_{3})$.
Directly encode $x^{n},y^{n}$ into $m_{1}$ and $m_{2}$ respectively.

Decoder 3 receives $m_{0},m_{3},y^{n}$ and output the unique $\hat{x}^{n}$
such that $(\hat{x}^{n},y^{n})\in\mathcal{B}_{3}(m_{0},m_{3})$. The
probability of error vanishes if $H(Y)>H(X,Y)+\epsilon/2-R_{0}-R_{3}$,
which is guaranteed by the definition of $R(p_{7})$. Decoder 4 receives
$m_{0},x^{n}$ and output the unique $\hat{y}^{n}$ such that $(x^{n},\hat{y}^{n})\in\mathcal{B}_{0}(m_{0})$.
The probability of error vanishes if $H(X)>H(X,Y)+\epsilon/2-R_{0}$,
which is guaranteed by the definition of $R(p_{7})$.\medskip{}

The achievability of $R(p_{8})$ is similar to that of $R(p_{7})$.
To generate the codebook, randomly partition $\mathcal{T}_{\epsilon'}^{(n)}(X,Y)$
into bins $\mathcal{B}_{0}(m_{0})$ of size $2^{n(H(X,Y)+\epsilon/2-R_{0})}$
for $m_{0}\in[1:\,2^{nR_{0}}]$. Given $m_{0}$, assign indices $m_{1}$
to the sequences in the bin $\mathcal{B}_{0}(m_{0})$ for $m_{1}\in[1:2^{nR_{1}}]$.
This is possible if $R_{1}\ge H(X,Y)+\epsilon/2-R_{0}$, which is
guaranteed by the definition of $R(p_{8})$.

To encode $x^{n},y^{n}$, find $m_{0}$ such that $(x^{n},y^{n})\in\mathcal{B}_{0}(m_{0})$
and find the index $m_{1}$. Directly encode $y^{n}$ into $m_{2}$.

Decoder 1 receives $m_{0},m_{1}$ and output $x^{n}$ where $(x^{n},y^{n})\in\mathcal{B}_{0}(m_{0})$
with index $m_{1}$. Decoder 3 receives $m_{0},y^{n}$ and output
the unique $\hat{x}^{n}$ such that $(\hat{x}^{n},y^{n})\in\mathcal{B}_{0}(m_{0})$.
The probability of error vanishes if $H(Y)>H(X,Y)+\epsilon/2-R_{0}$,
which is guaranteed by the definition of $R(p_{8})$. Decoder 4 receives
$m_{0},x^{n}$ and output the unique $\hat{y}^{n}$ such that $(x^{n},\hat{y}^{n})\in\mathcal{B}_{0}(m_{0})$.
The probability of error vanishes if $H(X)>H(X,Y)+\epsilon/2-R_{0}$,
which follows from the definition of $R(p_{8})$ and $H(X)\ge H(Y)$.\medskip{}

Hence we have proved that for any point $v\in\mathscr{I}_{XY}$ and
\[
w\in\mathscr{I}(v)=\left((-\infty,v_{X}]\times(-\infty,v_{Y}]\times[v_{XY},\infty)\right)\cap\mathscr{I}^{\mathrm{o}}_{XY},
\]
the rate tuple $R(w)$ is achievable. In other words, the region
\[
R\left(\left(\mathscr{I}_{XY}+(-\infty,0]\times(-\infty,0]\times[0,\infty)\right)\cap\mathscr{I}^{\mathrm{o}}_{XY}\right)+[0,\infty)^{5}
\]
is achievable. The region can be written as
\begin{align*}
w_{XY} & \ge I(X,Y;U),\\
w_{X} & \le I(X;U),\\
w_{Y} & \le I(Y;U),\\
w_{X} & \ge0,\\
w_{Y} & \ge0,\\
w_{XY}-w_{Y} & \le H(X|Y),\\
w_{XY}-w_{X} & \le H(Y|X),\\
R_{0} & \ge w_{XY}+\epsilon,\\
R_{1} & \ge H(X)-w_{X}+\epsilon,\\
R_{2} & \ge H(Y)-w_{Y}+\epsilon,\\
R_{3} & \ge H(X|Y)-w_{XY}+w_{Y}+\epsilon,\\
R_{4} & \ge H(Y|X)-w_{XY}+w_{X}+\epsilon
\end{align*}
for some $U,w_{X},w_{Y},w_{XY}$. The final rate region can be obtained
by eliminating $w_{X},w_{Y},w_{XY}$ using Fourier-Motzkin elimination.\medskip{}
\medskip{}

We then prove the converse. Since decoder 3 observes $M_{0},M_{3},Y^{n}$
and has to recover $X^{n}$ with vanishing error probability, $R_{0}+R_{3}\ge H(X|Y)$.
Similarly $R_{0}+R_{4}\ge H(Y|X)$. Note that decoders 2 and 3 together
can recover $X^{n},Y^{n}$ with vanishing error probability (decoder
3 uses the output of decoder 2 as the side information), and hence
$R_{0}+R_{2}+R_{3}\ge H(X,Y)$. Similarly $R_{0}+R_{1}+R_{4}\ge H(X,Y)$.

Let $U_{i}=(M_{0},X^{i-1},Y^{i-1})$. Using the same arguments in
the proof of Theorem \ref{thm:gwc}, we have $R_{0}\ge I(X,Y;U)$,
$R_{1}\ge H(X|U)$, $R_{2}\ge H(Y|U)$.
\begin{align*}
nR_{3} & \ge H(M_{3}|M_{0})\\
 & \ge I(X^{n};\,M_{3}|M_{0})\\
 & =H(X^{n}|M_{0})-H(X^{n}|M_{0},M_{3})\\
 & =\sum_{i=1}^{n}H(X_{i}|M_{0},X^{i-1})-H(X^{n}|M_{0},M_{3})\\
 & \ge\sum_{i=1}^{n}H(X_{i}|M_{0},X^{i-1},Y^{i-1})-H(Y^{n})-H(X^{n}|M_{0},M_{3},Y^{n})\\
 & \ge\sum_{i=1}^{n}H(X_{i}|U_{i})-H(Y^{n})-o(n),
\end{align*}
where the last inequality is due to Fano's inequality. Similarly $nR_{4}\ge\sum_{i}H(Y_{i}|U_{i})-H(X^{n})-o(n)$.
\begin{align*}
 & n(R_{2}+R_{3})\\
 & \ge H(M_{2},M_{3}|M_{0})\\
 & \ge I(X^{n};\,M_{2},M_{3}|M_{0})\\
 & =H(X^{n}|M_{0})-H(X^{n}|M_{0},M_{2},M_{3})\\
 & =\sum_{i=1}^{n}H(X_{i}|M_{0},X^{i-1})-H(X^{n}|M_{0},M_{2},M_{3})\\
 & \ge\sum_{i=1}^{n}H(X_{i}|M_{0},X^{i-1},Y^{i-1})-H(Y^{n}|M_{0},M_{2},M_{3})-H(X^{n}|M_{0},M_{2},M_{3},Y^{n})\\
 & \ge\sum_{i=1}^{n}H(X_{i}|U_{i})-o(n),
\end{align*}
where the last inequality follows by Fano's inequality. Similarly $n(R_{1}+R_{4})\ge\sum_{i}H(Y_{i}|U_{i})-o(n)$.
Hence the point $(R_{0}+\epsilon,\ldots,R_{4}+\epsilon)$ is in the
convex hull of $\Rr'$ for any $\epsilon>0$.
We have seen in the achievability proof that (for $\epsilon=0$)
\[
\Rr'=R\left(\left(\mathscr{I}_{XY}+(-\infty,0]\times(-\infty,0]\times[0,\infty)\right)\cap\mathscr{I}^{\mathrm{o}}_{XY}\right)+[0,\infty)^{5}
\]
is the increasing hull of an affine transformation of a convex set.
Therefore $\Rr'$ is convex.

\subsection{Proof of Proposition \ref{prop:cl_ixy}\label{subsec:pf_cl_ixy}}
\begin{enumerate}
\item Since the Gray--Wyner region tensorizes, $\mathrm{cl}(\mathscr{I}_{XY}^{\infty})\subseteq\left(\mathscr{I}_{XY}+(-\infty,0]\times(-\infty,0]\times[0,\infty)\right)\cap\mathscr{I}^{\mathrm{o}}_{XY}$.
To prove the other direction, let $w\in\left(\mathscr{I}_{XY}+(-\infty,0]\times(-\infty,0]\times[0,\infty)\right)\cap\mathscr{I}^{\mathrm{o}}_{XY}$,
then by Theorem~\ref{thm:gwn}, the following rate tuple is achievable
\begin{align*}
R_{0}(w) & =w_{XY}+\epsilon,\\
R_{1}(w) & =H(X)-w_{X}+\epsilon,\\
R_{2}(w) & =H(Y)-w_{Y}+\epsilon,\\
R_{3}(w) & =H(X|Y)-w_{XY}+w_{Y}+\epsilon,\\
R_{4}(w) & =H(Y|X)-w_{XY}+w_{X}+\epsilon,
\end{align*}
i.e. for the source $X^{l},Y^{l}$, the probability of error $P_{e}(l)\to0$
as $l\to\infty$. Apply this scheme $n$ times on the source $X^{nl},Y^{nl}$.
This can be considered as a causal scheme on the source sequence $(X_{1}^{l},Y_{1}^{l}),(X_{l+1}^{2l},Y_{l+1}^{2l}),\ldots,(X_{(n-1)l+1}^{nl},Y_{(n-1)l+1}^{nl})$
with rate tuple $lR(w)$ and symbol error probability $P_{e}(l)$.
Hence by \eqref{eq:gwc_symerr} and \eqref{eq:gwc_symerr2} in the
proof of Theorem \ref{thm:gwc}, 
\[
R(w)+\log\left(\left|\mathcal{X}\right|\cdot\left|\mathcal{Y}\right|\right)P_{e}(l)\cdot\mathbf{1}\in(1/l)\mathscr{R}(X^{l};Y^{l}).
\]
Let $\epsilon'=\epsilon+\log\left(\left|\mathcal{X}\right|\cdot\left|\mathcal{Y}\right|\right)P_{e}(l)$.
Since
\begin{align*}
\frac{1}{l}\mathscr{R}(X^{l};Y^{l})=\bigcup_{v\in(1/l)\mathscr{I}_{X^{l}Y^{l}}} & [v_{XY},\infty)\times[H(X)-v_{X},\infty)\times[H(Y)-v_{Y},\infty)\\
 & \times[H(X|Y)-v_{XY}+v_{Y},\infty)\times[H(Y|X)-v_{XY}+v_{X},\infty),
\end{align*}
there exists $v\in(1/l)\mathscr{I}_{X^{l}Y^{l}}\subseteq\mathscr{I}_{XY}^{\infty}$
such that $v_{XY}\le w_{XY}+\epsilon'$, $H(X)-v_{X}\le H(X)-w_{X}+\epsilon'$,
and similar for the other 3 dimensions, which implies $\left\Vert v-w\right\Vert _{\infty}\le2\epsilon'$.
The result follows from taking $l\to\infty$, $\epsilon\to0$.

To show 
\begin{align*}
& \left(\mathscr{I}_{XY}+(-\infty,0]\times(-\infty,0]\times[0,\infty)\right)\cap\mathscr{I}^{\mathrm{o}}_{XY}\\
&=\left(\mathscr{I}_{XY}+ \{(t, t, t) :\, t \le 0 \} \right)\cap \big( [0,\infty) \times [0,\infty) \times \mathbb{R} \big),
\end{align*}
note that they are both equal to the union of the convex hulls of $\{ v, p_1,\ldots,p_8 \}$ for $v \in \mathscr{I}_{XY}$ (as in the proof of Theorem~\ref{thm:gwn}).

\item The equivalence between $\mathrm{cl}(\mathscr{I}_{XY}^{\infty})$
and $\Rr'$ is proved in the Fourier-Motzkin elimination step in the
proof of Theorem \ref{thm:gwn}.
\item By Proposition \ref{prop:ixy_prop},
\begin{align*}
\mathscr{R}_{\mathrm{GW}} & =\bigcup_{v\in\mathscr{I}_{XY}}\big\{\big(v_{XY},\,H(X)-v_{X},\,H(Y)-v_{Y}\big)\big\}+[0,\infty)^{3}\\
 & =\bigcup_{v\in\mathscr{I}_{XY}\cap\mathscr{I}^{\mathrm{o}}_{XY}}\big\{\big(v_{XY},\,H(X)-v_{X},\,H(Y)-v_{Y}\big)\big\}+[0,\infty)^{3}\\
 & =\bigcup_{v\in\left(\mathscr{I}_{XY}+(-\infty,0]^{2}\times[0,\infty)\right)\cap\mathscr{I}^{\mathrm{o}}_{XY}}\big\{\big(v_{XY},\,H(X)-v_{X},\,H(Y)-v_{Y}\big)\big\}+[0,\infty)^{3}\\
 & =\bigcup_{v\in\mathrm{cl}(\mathscr{I}_{XY}^{\infty})}\big\{\big(v_{XY},\,H(X)-v_{X},\,H(Y)-v_{Y}\big)\big\}+[0,\infty)^{3}.
\end{align*}
For the other direction,
\begin{align*}
\mathrm{cl}(\mathscr{I}_{XY}^{\infty}) & =\left(\mathscr{I}_{XY}+(-\infty,0]^{2}\times[0,\infty)\right)\cap\mathscr{I}^{\mathrm{o}}_{XY}\\
 & =\left\{ v\in\mathscr{I}^{\mathrm{o}}_{XY}:\,v_{X}\le w_{X},\,v_{Y}\le w_{Y},\,v_{XY}\ge w_{XY}\;\text{for some}\;w\in\mathscr{I}_{XY}\right\} \\
 & =\left\{ v\in\mathscr{I}^{\mathrm{o}}_{XY}:\,v_{X}\le I(X;U),\,v_{Y}\le I(Y;U),\,v_{XY}\ge I(X,Y;U)\;\text{for some}\;U\right\} \\
 & =\left\{ v\in\mathscr{I}^{\mathrm{o}}_{XY}:\,H(X)-v_{X}\ge H(X|U),\,H(Y)-v_{Y}\ge H(Y|U),\,v_{XY}\ge I(X,Y;U)\;\text{for some}\;U\right\} \\
 & =\left\{ v\in\mathscr{I}^{\mathrm{o}}_{XY}:\,\big(v_{XY},\,H(X)-v_{X},\,H(Y)-v_{Y}\big)\in\mathscr{R}_{\mathrm{GW}}\right\} .
\end{align*}
\end{enumerate}

\bibliographystyle{IEEEtran}
\bibliography{ref,nit}

\newcommand{\noopsort}[1]{}
\begin{thebibliography}{10}
\providecommand{\url}[1]{#1}
\csname url@samestyle\endcsname
\providecommand{\newblock}{\relax}
\providecommand{\bibinfo}[2]{#2}
\providecommand{\BIBentrySTDinterwordspacing}{\spaceskip=0pt\relax}
\providecommand{\BIBentryALTinterwordstretchfactor}{4}
\providecommand{\BIBentryALTinterwordspacing}{\spaceskip=\fontdimen2\font plus
\BIBentryALTinterwordstretchfactor\fontdimen3\font minus
  \fontdimen4\font\relax}
\providecommand{\BIBforeignlanguage}[2]{{%
\expandafter\ifx\csname l@#1\endcsname\relax
\typeout{** WARNING: IEEEtran.bst: No hyphenation pattern has been}%
\typeout{** loaded for the language `#1'. Using the pattern for}%
\typeout{** the default language instead.}%
\else
\language=\csname l@#1\endcsname
\fi
#2}}
\providecommand{\BIBdecl}{\relax}
\BIBdecl

\bibitem{Gray--Wyner1974}
R.~M. Gray and A.~D. Wyner, ``Source coding for a simple network,'' \emph{Bell
  Syst. Tech. J.}, vol.~53, no.~9, pp. 1681--1721, 1974.

\bibitem{Wyner1975a}
A.~D. Wyner, ``The common information of two dependent random variables,''
  \emph{{IEEE} Trans. Inf. Theory}, vol.~21, no.~2, pp. 163--179, Mar. 1975.

\bibitem{Gacs--Korner1973}
P.~G{\'a}cs and J.~K{\"o}rner, ``Common information is far less than mutual
  information,'' \emph{Probl. Control Inf. Theory}, vol.~2, no.~2, pp.
  149--162, 1973.

\bibitem{cuff2010coordination}
P.~Cuff, H.~Permuter, and T.~M. Cover, ``Coordination capacity,'' \emph{IEEE
  Trans. Info. Theory}, vol.~56, no.~9, pp. 4181--4206, 2010.

\bibitem{tishby2000information}
N.~Tishby, F.~C. Pereira, and W.~Bialek, ``The information bottleneck method,''
  \emph{arXiv preprint physics/0004057}, 2000.

\bibitem{wyner2002satellite}
A.~D. Wyner, J.~K. Wolf, and F.~M.~J. Willems, ``Communicating via a processing
  broadcast satellite,'' \emph{IEEE Transactions on Information Theory},
  vol.~48, no.~6, pp. 1243--1249, Jun 2002.

\bibitem{Kimura--Uyematsu2006}
A.~Kimura and T.~Uyematsu, ``Multiterminal source coding with complementary
  delivery,'' in \emph{Proc. {IEEE} Int. Symp. Inf. Theory Appl.}, Seoul,
  Korea, October 2006, pp. 189--194.

\bibitem{korner1973coding}
J.~K{\"o}rner, ``Coding of an information source having ambiguous alphabet and
  the entropy of graphs,'' in \emph{6th Prague conference on information
  theory}, 1973, pp. 411--425.

\bibitem{makhdoumi2014funnel}
A.~Makhdoumi, S.~Salamatian, N.~Fawaz, and M.~Medard, ``From the information
  bottleneck to the privacy funnel,'' in \emph{Information Theory Workshop
  (ITW), 2014 IEEE}, Nov 2014, pp. 501--505.

\bibitem{sfrl_arxiv}
\BIBentryALTinterwordspacing
C.~T. Li and A.~{El Gamal}, ``Strong functional representation lemma and
  applications to coding theorems,'' \emph{arXiv preprint}, 2017. [Online].
  Available: \url{http://arxiv.org/abs/1701.02827}
\BIBentrySTDinterwordspacing

\bibitem{mcgill1954multivariate}
\BIBentryALTinterwordspacing
W.~J. McGill, ``Multivariate information transmission,'' \emph{Psychometrika},
  vol.~19, no.~2, pp. 97--116, 1954. [Online]. Available:
  \url{http://dx.doi.org/10.1007/BF02289159}
\BIBentrySTDinterwordspacing

\bibitem{Weissman--El-Gamal2006}
T.~Weissman and A.~El~Gamal, ``Source coding with limited-look-ahead side
  information at the decoder,'' \emph{{IEEE} Trans. Inf. Theory}, vol.~52,
  no.~12, pp. 5218--5239, Dec. 2006.

\bibitem{cuff2013distributed}
P.~Cuff, ``Distributed channel synthesis,'' \emph{IEEE Trans. Info. Theory},
  vol.~59, no.~11, pp. 7071--7096, 2013.

\bibitem{bennet2014reverse}
C.~H. Bennett, I.~Devetak, A.~W. Harrow, P.~W. Shor, and A.~Winter, ``The
  quantum reverse shannon theorem and resource tradeoffs for simulating quantum
  channels,'' \emph{IEEE Trans. Info. Theory}, vol.~60, no.~5, pp. 2926--2959,
  May 2014.

\bibitem{Wyner1973b}
A.~D. Wyner, ``A theorem on the entropy of certain binary sequences and
  applications---{II},'' \emph{{IEEE} Trans. Inf. Theory}, vol.~19, no.~6, pp.
  772--777, 1973.

\bibitem{Ahlswede--Korner1975}
R.~Ahlswede and J.~K{\"o}rner, ``Source coding with side information and a
  converse for degraded broadcast channels,'' \emph{{IEEE} Trans. Inf. Theory},
  vol.~21, no.~6, pp. 629--637, 1975.

\bibitem{Wyner1975b}
A.~D. Wyner, ``On source coding with side information at the decoder,''
  \emph{{IEEE} Trans. Inf. Theory}, vol.~21, no.~3, pp. 294--300, 1975.

\bibitem{timo2009causal}
R.~Timo and B.~N. Vellambi, ``Two lossy source coding problems with causal
  side-information,'' in \emph{2009 IEEE International Symposium on Information
  Theory}, June 2009, pp. 1040--1044.

\bibitem{Wyner--Ziv1976}
A.~D. Wyner and J.~Ziv, ``The rate--distortion function for source coding with
  side information at the decoder,'' \emph{{IEEE} Trans. Inf. Theory}, vol.~22,
  no.~1, pp. 1--10, 1976.

\bibitem{heegard1985sideinfo}
C.~Heegard and T.~Berger, ``Rate distortion when side information may be
  absent,'' \emph{IEEE Transactions on Information Theory}, vol.~31, no.~6, pp.
  727--734, Nov 1985.

\bibitem{steinberg2004successive}
Y.~Steinberg and N.~Merhav, ``On successive refinement for the {W}yner-{Z}iv
  problem,'' in \emph{International Symposium onInformation Theory, 2004. ISIT
  2004. Proceedings.}, 2004, pp. 364--364.

\bibitem{tian2008sideinfo}
C.~Tian and S.~N. Diggavi, ``Side-information scalable source coding,''
  \emph{IEEE Transactions on Information Theory}, vol.~54, no.~12, pp.
  5591--5608, Dec 2008.

\bibitem{prabhakaran2014assisted}
V.~M. Prabhakaran and M.~M. Prabhakaran, ``Assisted common information with an
  application to secure two-party sampling,'' \emph{IEEE Transactions on
  Information Theory}, vol.~60, no.~6, pp. 3413--3434, June 2014.

\bibitem{prabhakaran2014tension}
M.~M. Prabhakaran and V.~M. Prabhakaran, ``Tension bounds for information
  complexity,'' \emph{arXiv preprint arXiv:1408.6285}, 2014.

\bibitem{witsenhausen1975sequences}
H.~S. Witsenhausen, ``On sequences of pairs of dependent random variables,''
  \emph{SIAM Journal on Applied Mathematics}, vol.~28, no.~1, pp. 100--113,
  1975.

\bibitem{alon1996}
N.~Alon and A.~Orlitsky, ``Source coding and graph entropies,'' \emph{IEEE
  Transactions on Information Theory}, vol.~42, no.~5, pp. 1329--1339, Sep
  1996.

\bibitem{wolf2008monotones}
S.~Wolf and J.~Wullschleger, ``New monotones and lower bounds in unconditional
  two-party computation,'' \emph{IEEE Transactions on Information Theory},
  vol.~54, no.~6, pp. 2792--2797, June 2008.

\bibitem{kamath2010}
S.~Kamath and V.~Anantharam, ``A new dual to the {G}{\'a}cs-{K}{\"o}rner common
  information defined via the {G}ray-{W}yner system,'' in \emph{Communication,
  Control, and Computing (Allerton), 2010 48th Annual Allerton Conference on},
  Sept 2010, pp. 1340--1346.

\bibitem{asoodeh2014privacy}
S.~Asoodeh, F.~Alajaji, and T.~Linder, ``Notes on information-theoretic
  privacy,'' in \emph{Communication, Control, and Computing (Allerton), 2014
  52nd Annual Allerton Conference on}, Sept 2014, pp. 1272--1278.

\bibitem{banerjee2015synergy}
P.~K. Banerjee and V.~Griffith, ``Synergy, redundancy and common information,''
  \emph{arXiv preprint arXiv:1509.03706}, 2015.

\bibitem{ahlswede1998cr}
R.~Ahlswede and I.~Csiszar, ``Common randomness in information theory and
  cryptography. {II.} {CR} capacity,'' \emph{IEEE Transactions on Information
  Theory}, vol.~44, no.~1, pp. 225--240, Jan 1998.

\bibitem{witsenhausen1975conditional}
H.~Witsenhausen and A.~Wyner, ``A conditional entropy bound for a pair of
  discrete random variables,'' \emph{IEEE Transactions on Information Theory},
  vol.~21, no.~5, pp. 493--501, Sep 1975.

\bibitem{Anantharam2013maxcor}
\BIBentryALTinterwordspacing
V.~Anantharam, A.~A. Gohari, S.~Kamath, and C.~Nair, ``On maximal correlation,
  hypercontractivity, and the data processing inequality studied by erkip and
  cover,'' \emph{CoRR}, vol. abs/1304.6133, 2013. [Online]. Available:
  \url{http://arxiv.org/abs/1304.6133}
\BIBentrySTDinterwordspacing

\bibitem{ahlswede1976spreading}
R.~Ahlswede and P.~G{\'a}cs, ``Spreading of sets in product spaces and
  hypercontraction of the {M}arkov operator,'' \emph{The annals of
  probability}, pp. 925--939, 1976.

\bibitem{calmon2015perfect}
F.~P. Calmon, A.~Makhdoumi, and M.~Medard, ``Fundamental limits of perfect
  privacy,'' in \emph{2015 IEEE International Symposium on Information Theory
  (ISIT)}, June 2015, pp. 1796--1800.

\bibitem{harsha2010communication}
P.~Harsha, R.~Jain, D.~McAllester, and J.~Radhakrishnan, ``The communication
  complexity of correlation,'' \emph{IEEE Trans. Info. Theory}, vol.~56, no.~1,
  pp. 438--449, Jan 2010.

\bibitem{knuth1976complexity}
D.~E. Knuth and A.~C. Yao, ``The complexity of nonuniform random number
  generation,'' \emph{Algorithms and complexity: new directions and recent
  results}, pp. 357--428, 1976.

\bibitem{Eggleston1958}
H.~G. Eggleston, \emph{Convexity}.\hskip 1em plus 0.5em minus 0.4em\relax
  Cambridge: Cambridge University Press, 1958.

\bibitem{Rockafellar1970}
R.~T. Rockafellar, \emph{Convex Analysis}.\hskip 1em plus 0.5em minus
  0.4em\relax Princeton, NJ: Princeton University Press, 1970.

\bibitem{elgamal2011network}
A.~El~Gamal and Y.-H. Kim, \emph{Network information theory}.\hskip 1em plus
  0.5em minus 0.4em\relax Cambridge university press, 2011.

\end{thebibliography}

\end{document}